\documentclass[twocolumn]{openjournal}
\usepackage{enumitem}
\usepackage{bm}
\usepackage{booktabs}

\usepackage{tikz}
\usetikzlibrary{trees}
\usepackage{amsmath,amsfonts,amssymb}
\usepackage{graphicx}
\usepackage{listings}
\usepackage{txfonts}
\usepackage{color}
\usepackage{textcomp}
\usepackage{natbib}
\usepackage{import}
\usepackage{float}
\usepackage{afterpage}
\usepackage{ifthen}
\usepackage[morefloats=12]{morefloats}
\usepackage{placeins}
\bibpunct{(}{)}{;}{a}{}{,}
\usepackage[switch]{lineno}
\PassOptionsToPackage{hyphens}{url}
\usepackage{url}

\definecolor{linkcolor}{rgb}{0.6,0,0}
\definecolor{citecolor}{rgb}{0,0,0.75}
\definecolor{urlcolor}{rgb}{0.12,0.46,0.7}
\usepackage[breaklinks, colorlinks, urlcolor=urlcolor,
    linkcolor=linkcolor,citecolor=citecolor,pdfencoding=auto]{hyperref}
\hypersetup{linktocpage}
\usepackage{bold-extra}
\usepackage{ragged2e}

\def\setsymbol#1#2{\expandafter\def\csname #1\endcsname{#2}}
\def\getsymbol#1{\csname #1\endcsname}

\def\Planck{\textit{Planck}}

\newbox\tablebox    \newdimen\tablewidth
\def\leaderfil{\leaders\hbox to 5pt{\hss.\hss}\hfil}
\def\tablenote#1 #2\par{\begingroup \parindent=0.8em
    \abovedisplayshortskip=0pt\belowdisplayshortskip=0pt
    \noindent
    $$\hss\vbox{\hsize\tablewidth \hangindent=\parindent \hangafter=1 \noindent
    \hbox to \parindent{$^#1$\hss}\strut#2\strut\par}\hss$$
    \endgroup}

\def\L2{\ifmmode L_2\else $L_2$\fi}

\def\DeltaT{\ifmmode \Delta T\else $\Delta T$\fi}
\def\deltat{\ifmmode \Delta t\else $\Delta t$\fi}
\def\fknee{\ifmmode f_{\rm knee}\else $f_{\rm knee}$\fi}
\def\Fmax{\ifmmode F_{\rm max}\else $F_{\rm max}$\fi}
\def\solar{\ifmmode{\rm M}_{\mathord\odot}\else${\rm M}_{\mathord\odot}$\fi}
\def\Msolar{\ifmmode{\rm M}_{\mathord\odot}\else${\rm M}_{\mathord\odot}$\fi}
\def\Lsolar{\ifmmode{\rm L}_{\mathord\odot}\else${\rm L}_{\mathord\odot}$\fi}
\def\inv{\ifmmode^{-1}\else$^{-1}$\fi}
\def\mo{\ifmmode^{-1}\else$^{-1}$\fi}
\def\sup#1{\ifmmode ^{\rm #1}\else $^{\rm #1}$\fi}
\def\expo#1{\ifmmode \times 10^{#1}\else $\times 10^{#1}$\fi}
\def\,{\thinspace}
\def\lsim{\mathrel{\raise .4ex\hbox{\rlap{$<$}\lower 1.2ex\hbox{$\sim$}}}}
\def\gsim{\mathrel{\raise .4ex\hbox{\rlap{$>$}\lower 1.2ex\hbox{$\sim$}}}}

\def\simprop{\mathrel{\raise .4ex\hbox{\rlap{$\propto$}\lower 1.2ex\hbox{$\sim$}}}}
\def\deg{\ifmmode^\circ\else$^\circ$\fi}
\def\pdeg{\ifmmode $\setbox0=\hbox{$^{\circ}$}\rlap{\hskip.11\wd0 .}$^{\circ}
          \else \setbox0=\hbox{$^{\circ}$}\rlap{\hskip.11\wd0 .}$^{\circ}$\fi}
\def\arcs{\ifmmode {^{\scriptstyle\prime\prime}}
          \else $^{\scriptstyle\prime\prime}$\fi}
\def\arcm{\ifmmode {^{\scriptstyle\prime}}
          \else $^{\scriptstyle\prime}$\fi}
\newdimen\sa  \newdimen\sb
\def\parcs{\sa=.07em \sb=.03em
     \ifmmode \hbox{\rlap{.}}^{\scriptstyle\prime\kern -\sb\prime}\hbox{\kern -\sa}
     \else \rlap{.}$^{\scriptstyle\prime\kern -\sb\prime}$\kern -\sa\fi}
\def\parcm{\sa=.08em \sb=.03em
     \ifmmode \hbox{\rlap{.}\kern\sa}^{\scriptstyle\prime}\hbox{\kern-\sb}
     \else \rlap{.}\kern\sa$^{\scriptstyle\prime}$\kern-\sb\fi}
\def\ra[#1 #2 #3.#4]{#1\sup{h}#2\sup{m}#3\sup{s}\llap.#4}
\def\dec[#1 #2 #3.#4]{#1\deg#2\arcm#3\arcs\llap.#4}
\def\deco[#1 #2 #3]{#1\deg#2\arcm#3\arcs}
\def\rra[#1 #2]{#1\sup{h}#2\sup{m}}

\def\dots{\relax\ifmmode \ldots\else $\ldots$\fi}
\def\WHzsr{\ifmmode $W\,Hz\mo\,sr\mo$\else W\,Hz\mo\,sr\mo\fi}
\def\mHz{\ifmmode $\,mHz$\else \,mHz\fi}
\def\GHz{\ifmmode $\,GHz$\else \,GHz\fi}
\def\mKs{\ifmmode $\,mK\,s$^{1/2}\else \,mK\,s$^{1/2}$\fi}
\def\muKs{\ifmmode \,\mu$K\,s$^{1/2}\else \,$\mu$K\,s$^{1/2}$\fi}
\def\muKRJs{\ifmmode \,\mu$K$_{\rm RJ}$\,s$^{1/2}\else \,$\mu$K$_{\rm RJ}$\,s$^{1/2}$\fi}
\def\muKHz{\ifmmode \,\mu$K\,Hz$^{-1/2}\else \,$\mu$K\,Hz$^{-1/2}$\fi}
\def\MJysr{\ifmmode \,$MJy\,sr\mo$\else \,MJy\,sr\mo\fi}
\def\MJysrmK{\ifmmode \,$MJy\,sr\mo$\,mK$_{\rm CMB}\mo\else \,MJy\,sr\mo\,mK$_{\rm CMB}\mo$\fi}
\def\microns{\ifmmode \,\mu$m$\else \,$\mu$m\fi}

\def\muK{\ifmmode \,\mu$K$\else \,$\mu$\hbox{K}\fi}
\def\microK{\ifmmode \,\mu$K$\else \,$\mu$\hbox{K}\fi}
\def\muW{\ifmmode \,\mu$W$\else \,$\mu$\hbox{W}\fi}
\def\kms{\ifmmode $\,km\,s$^{-1}\else \,km\,s$^{-1}$\fi}
\def\kmsMpc{\ifmmode $\,\kms\,Mpc\mo$\else \,\kms\,Mpc\mo\fi}
\providecommand{\sorthelp}[1]{}

\def\WMAP{\textit{WMAP}}
\def\COBE{\textit{COBE}}

\def\healpix{\texttt{HEALPix}}
\def\commander{\texttt{Commander}}

\def\commanderthree{\texttt{Commander3}}

\renewcommand{\d}[0]{\vec{d}}

\newcommand{\n}[0]{\vec{n}}

\newcommand{\s}[0]{\vec{s}}

\renewcommand{\L}[0]{\tens{L}}

\newcommand{\BP}{\textsc{BeyondPlanck}}
\newcommand{\cosmoglobe}{\textsc{Cosmoglobe}}

\DeclareTextFontCommand{\textbfit}{%
  \fontseries\bfdefault 
  \itshape
}

\def\inv{^{-1}}

\begin{document}

\title{\bfseries{From \scshape{BeyondPlanck} to \scshape{Cosmoglobe}:}\\
    Open Science, Reproducibility, and Data Longevity}

\newcommand{\planetek}[0]{$^{1}$}
\newcommand{\oslo}[0]{$^{2}$}
\newcommand{\milanoA}[0]{$^{3}$}
\newcommand{\milanoB}[0]{$^{4}$}
\newcommand{\milanoC}[0]{$^{5}$}
\newcommand{\triesteB}[0]{$^{6}$}
\newcommand{\princeton}[0]{$^{7}$}
\newcommand{\jpl}[0]{$^{8}$}
\newcommand{\helsinkiA}[0]{$^{9}$}
\newcommand{\helsinkiB}[0]{$^{10}$}
\newcommand{\nersc}[0]{$^{11}$}
\newcommand{\mpa}[0]{$^{12}$}
\newcommand{\triesteA}[0]{$^{13}$}
\newcommand{\haverford}[0]{$^{14}$}

\author{\small
S.~Gerakakis\planetek,
M.~Brilenkov\oslo$^{*}$,
M.~Ieronymaki\planetek,
M.~San{\oslo},
D.~J.~Watts{\oslo},
\and
K.~J.~Andersen{\oslo},
\and
\textcolor{black}{R.~Aurlien}{\oslo},
\and
\textcolor{black}{R.~Banerji}{\oslo},
\and
\textcolor{black}{A.~Basyrov}{\oslo},
\and
M.~Bersanelli{\milanoA}$^{,}${\milanoB}$^{,}${\milanoC},
\and
S.~Bertocco{\triesteB},
\and
M.~Carbone{\planetek},
\and
L.~P.~L.~Colombo{\milanoA},
\and
H.~K.~Eriksen{\oslo},
\and
J.~R.~Eskilt{\oslo},
\and
\textcolor{black}{M.~K.~Foss}{\oslo},
\and
C.~Franceschet{\milanoA}$^{,}${\milanoC},
\and
\textcolor{black}{U.~Fuskeland}{\oslo},
\and
S.~Galeotta{\triesteB},
\and
M.~Galloway{\oslo},
\and
E.~Gjerl{\o}w{\oslo},
\and
\textcolor{black}{B.~Hensley}{\princeton},
\and
\textcolor{black}{D.~Herman}{\oslo},
\and
M.~Iacobellis{\planetek},
\and
\textcolor{black}{H.~T.~Ihle}{\oslo},
\and
J.~B.~Jewell{\jpl},
\and
\textcolor{black}{A.~Karakci}{\oslo},
\and
E.~Keih\"{a}nen{\helsinkiA}$^{,}${\helsinkiB},
\and
R.~Keskitalo{\nersc},
\and
J.~G.~S.~Lunde{\oslo},
\and
G.~Maggio{\triesteB},
\and
D.~Maino{\milanoA}$^{,}${\milanoB}$^{,}${\milanoC},
\and
M.~Maris{\triesteB},
\and
S.~Paradiso{\milanoA}$^{,}${\milanoB},
\and
M.~Reinecke{\mpa},
\and
N.-O.~Stutzer{\oslo},
\and
A.-S.~Suur-Uski{\helsinkiA}$^{,}${\helsinkiB},
\and
T.~L.~Svalheim{\oslo},
\and
D.~Tavagnacco{\triesteB}$^{,}${\triesteA},
\and
H.~Thommesen{\oslo},
\and
I.~K.~Wehus{\oslo},
\and
A.~Zacchei{\triesteB}
}
\affiliation{\small
{\planetek}Planetek Hellas, Leoforos Kifisias 44, Marousi 151 25, Greece\goodbreak
\and
{\oslo}Institute of Theoretical Astrophysics, University of Oslo, Blindern, Oslo, Norway\goodbreak
\and
{\milanoA}Dipartimento di Fisica, Universit\`{a} degli Studi di Milano, Via Celoria, 16, Milano, Italy\goodbreak
\and
{\milanoB}INAF-IASF Milano, Via E. Bassini 15, Milano, Italy\goodbreak
\and
{\milanoC}INFN, Sezione di Milano, Via Celoria 16, Milano, Italy\goodbreak
\and
{\triesteB}INAF - Osservatorio Astronomico di Trieste, Via G.B. Tiepolo 11, Trieste, Italy\goodbreak
\and
{\princeton}Department of Astrophysical Sciences, Princeton University, Princeton, NJ 08544,
U.S.A.\goodbreak
\and
{\jpl}Jet Propulsion Laboratory, California Institute of Technology, 4800 Oak Grove Drive, Pasadena, California, U.S.A.\goodbreak
\and
{\helsinkiA}Department of Physics, Gustaf H\"{a}llstr\"{o}min katu 2, University of Helsinki, Helsinki, Finland\goodbreak
\and
{\helsinkiB}Helsinki Institute of Physics, Gustaf H\"{a}llstr\"{o}min katu 2, University of Helsinki, Helsinki, Finland\goodbreak
\and
{\nersc}Computational Cosmology Center, Lawrence Berkeley National Laboratory, Berkeley, California, U.S.A.\goodbreak
\and
{\mpa}Max-Planck-Institut f\"{u}r Astrophysik, Karl-Schwarzschild-Str. 1, 85741 Garching, Germany\goodbreak
\and
{\triesteA}Dipartimento di Fisica, Universit\`{a} degli Studi di Trieste, via A. Valerio 2, Trieste, Italy\goodbreak
}
\email{$^*$maksym.brilenkov@astro.uio.no}

\begin{abstract}
  The \BP\ and \cosmoglobe\ collaborations have 
  implemented the first integrated Bayesian end-to-end analysis 
  pipeline for CMB experiments. The primary long-term motivation 
  for this work is to develop a common analysis platform that 
  supports efficient global joint analysis of complementary 
  radio, microwave, and sub-millimeter experiments. A strict 
  prerequisite for this to succeed is broad participation 
  from the CMB community, and two foundational aspects of the 
  program are therefore reproducibility and Open Science. 
  In this paper, we discuss our efforts toward this aim. We 
  also discuss measures toward facilitating easy code and data 
  distribution, community-based code documentation, user-friendly 
  compilation procedures, etc. This work represents the first 
  publicly released end-to-end CMB analysis pipeline that 
  includes raw data, source code, parameter files, and 
  documentation. We argue that such a complete pipeline 
  release should be a requirement for all major 
  future and publicly-funded CMB experiments, noting that 
  a full public release significantly increases data 
  longevity by ensuring that the data quality can be 
  improved whenever better processing techniques, complementary 
  datasets, or more computing power become available, and 
  thereby also taxpayers' value for money; providing only 
  raw data and final products is not sufficient to guarantee 
  full reproducibility in the future.
\end{abstract}



\section{Introduction}
\label{sec:introduction}

Reproducibility and replicability are two of the defining features of modern
science. Within the field of CMB cosmology, this has most typically been
realized in the form of competition between different experiments, each trying
to measure the same sky signal but with different instrumentation and analysis
techniques.\footnote{See, e.g.,
\url{https://en.wikipedia.org/wiki/List_of_cosmic_microwave_background_experiments}
for a list of previous, current, and future CMB experiments.} This approach has
been tremendously successful and has led to a cosmological concordance
$\Lambda$CDM model that is able to statistically describe nearly all currently
available cosmological observables with only six free parameters
\citep{planck2016-l06}.

The next major milestone for the CMB field is the potential detection of
primordial gravitational waves and large-scale B-mode polarization
\citep[e.g.,][]{kamionkowski:2016}. If successful, this measurement will have
far-reaching implications for our understanding of physics at ultra-high energy
scales and the creation of the universe. However, this is also an extremely
technologically challenging measurement because of the very faint expected
signal amplitude. According to current theories and limits, it is anticipated
to account for no more than a few tens of nanokelvin fluctuations on large
angular scales, which is to be compared with the amplitude of the CMB solar
dipole of 3.4\,mK \citep{fixsen2009}, and with polarized astrophysical
foreground contamination of tens of microkelvins
\citep[e.g.,][]{planck2016-l04}. A robust detection will therefore require a
relative instrumental calibration better than $\mathcal{O}(10^{-5})$ and
foreground suppression better than two orders of magnitude
\citep[e.g.,][]{bp07,bp14}.

As discussed by \citet{bp01}, this challenge imposes substantial requirements
in terms of analysis and modeling techniques. Most notably, because of the
intimate relationship between instrument calibration and astrophysical
component separation, it is very likely that the associated parameters must be
explored jointly, and it is also quite possible that data from different
sources and experiments must be analyzed jointly to break internal degeneracies
that exist within each experiment separately. As a concrete example, despite
having almost one hundred times as many detectors as \Planck\
\citep{planck2016-l01} and more than one order of magnitude higher map-level
large-scale polarization sensitivity, \textit{LiteBIRD}'s \citep{litebird2022}
intensity sensitivity will not match \Planck's, and the ultimate
\textit{LiteBIRD} data analysis will therefore undoubtedly directly involve
\Planck\ measurements.

There is every reason to expect this to hold true for virtually all current and
planned CMB experiments. The data from these experiments will benefit
significantly from, if not depend on, a joint analysis with other datasets
within iterative pipelines. Such an approach will maximize the amount of
secondary science extracted from the datasets and allow them to achieve their
primary science goals. Without exception, every single CMB experiment fielded
to date has had parameters to which it was not sufficiently sensitive on its
own, whether due to its observation strategy, detector design, or frequency
coverage. It has typically required massive efforts to devise algorithmic
priors or tricks to self-consistently mitigate these ``blind spots'' or
``poorly measured modes''. However, the optimal solution to solving such
problems is, of course, by combining datasets with \emph{different} blind
spots, such that one experiment can break the degeneracies observed by another.
One concrete example of this is the current
\BP\footnote{\url{https://beyondplanck.science}} analysis, which re-analyzes
the \Planck\ LFI observations within a Bayesian end-to-end framework, and uses
\WMAP\ data to break important degeneracies between large-scale CMB
polarization modes and the LFI gain \citep{bp01,bp07}. Conversely, the ongoing
\WMAP\ re-analysis by \citet{bp17} will hopefully be able to constrain \WMAP's
transmission imbalance parameters using information from \Planck, and, if
successful, this will improve the data quality of both experiments. Similarly,
once \textit{LiteBIRD} data become available, both \WMAP\ and \Planck\ should
be re-analyzed from scratch, exploiting the \textit{LiteBIRD}'s
state-of-the-art large-scale polarization information to further improve the
gain models of both experiments. In general, we therefore argue that for this
type of joint analysis to be possible, it is critically important for all
involved experiments to provide both raw data and a fully operational data
analysis pipeline that can be re-run by external scientists.

A main goal of \cosmoglobe\footnote{\url{https://cosmoglobe.uio.no}} is to establish a common platform for this type of joint analysis\footnote{Currently, the parallelization of \commander\ is the most effective if applied to relatively small, TB-sized data sets. In this case, 1 TB of raw TOD data requires $\sim O(10^5)$ CPU-hours. For a thorough discussion of the topic, please see \cite{bp03}.} that can process low-level uncalibrated CMB time-ordered data (TOD) from different sources directly into high-level astrophysical component maps and cosmological parameters. The first application of this work is a full re-analysis of the \Planck\ LFI observations \citep[][and references therein]{bp01}, while extensions to \textit{WMAP} \citep{bp17}, \textit{LiteBIRD} \citep{litebird2022}, SPIDER \citep{spider21}, \COBE-DIRBE \citep{hauser:1998} and others are on-going. However, for this work to be successful as a community-wide enterprise, it is necessary for all researchers to be able to reproduce the existing work, and integrate their own datasets into the analysis. As such, Open Science and reproducibility plays a critical role in this program.

In this paper, we summarize our efforts on reproducibility within the context
of \BP\ and \cosmoglobe. Its main goals are two-fold. First, it outlines the
Open Source implementation of these projects and represents a valuable starting
point for other experiments aiming to contribute to and build on this
framework. Second, we hope that this paper may serve as a reference for any
other future astrophysics and cosmology collaboration that wants to perform its
work in an Open Source setting; the issues and tasks that need to be addressed
in the context of \BP\ and \cosmoglobe\ are very likely to be similar for any
other project of a similar type. As such, a significant fraction of this paper
is spent on surveys of tools and topics that were explored during the initial
phases of the project but not ultimately chosen simply because this material
may be helpful for other collaborations.

It is important to emphasize that Open Source initiatives are common in astronomy \& astrophysical settings. For example, the Astropy Project\footnote{\url{https://www.astropy.org/index.html}} --- the community-driven initiative with the primary goal of developing the core astronomy package in \texttt{Python}. Others include but are not limited to OpenAstronomy\footnote{\url{https://openastronomy.org/}}, Cosmo-Statistics\footnote{\color{urlcolor}{\texttt{https://cosmostatistics-initiative.org/}}}, Deep Skies\footnote{\url{https://deepskieslab.com/}}, and Dark Machines\footnote{\url{https://darkmachines.org/}}. Furthermore, a number of astrophysical codes and libraries were developed throughout the years, among which the particularly important for CMB are \texttt{CAMB}\footnote{\url{https://github.com/cmbant/camb}} \citep{camb1,camb2,camb3,camb4,camb5}, \texttt{CosmoMC}\footnote{\url{https://cosmologist.info/cosmomc/}} \citep{cosmomc,cosmomc2}, \texttt{TOAST}\footnote{\url{https://github.com/hpc4cmb/toast}}, \texttt{FGBuster}\footnote{\url{https://github.com/fgbuster/fgbuster}} \citep{2009MNRAS.392..216S, PhysRevD.84.069907, PhysRevD.94.083526}, \texttt{healpy}\footnote{\url{https://github.com/healpy/healpy}} \citep{zonca2019}, and many others, most of which have convenient \texttt{Python} wrappers to ease the usage. In addition, the reproduction of \Planck\ 2018 power spectra was done by \cite{zack2021_a} using the Open Source \texttt{PSpipe}\footnote{\url{https://github.com/simonsobs/PSpipe}} package.

The rest of the paper is organized as follows: Starting in Sect.~\ref{sec:bp},
we briefly review the statistical framework used by \BP\ and \cosmoglobe, and
we discuss why we believe that these issues will only become increasingly
important for all future major cosmology and astrophysical missions. Next, in
Sect.~\ref{sec:reproducibility} we give an overview of various possible
productivity tools that might be useful for future experiments, as well as
different Open Source licenses. In Sect.~\ref{sec:compilation_support} we
provide an overview of the compilation support facilities implemented for the
current software, while in Sect.~\ref{sec:bp_reproducibility} we summarize our
documentation and accessibility efforts. Finally, we conclude in
Sect.~\ref{sec:conclusions}.

\section{\BP, \cosmoglobe, and data longevity}
\label{sec:bp}

\begin{figure}[t]
    \center
    \includegraphics[width=\linewidth]{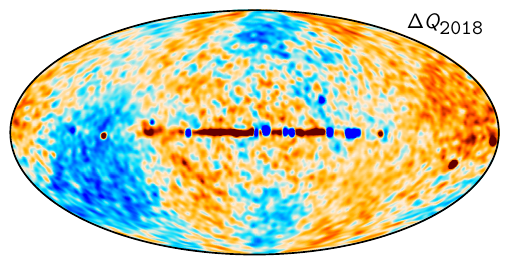}\\
    \includegraphics[width=\linewidth]{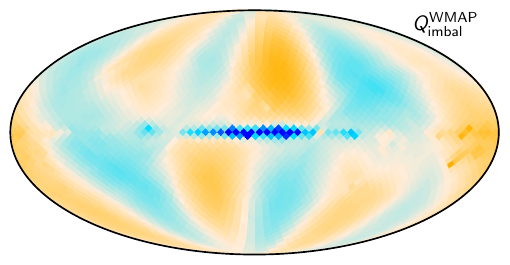}\\
    \includegraphics[width=\linewidth]{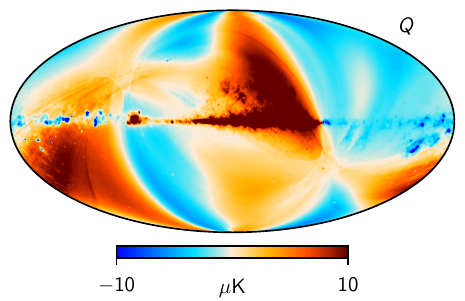}
    \caption{(\emph{Top row}:) Stokes $Q$ difference map between the 30\,GHz \Planck\ 2018 map and the $K$-band 9-year \WMAP\ map, smoothed to a common resolution of $3^{\circ}$ FWHM, and the latter has been scaled by a factor of 0.495 to account for different center frequencies; see \citet{bp07} for further discussion. (\emph{Middle row}:) \WMAP\ transmission imbalance template \citep{jarosik2007}. (\emph{Bottom row}:) \Planck\ 30\,GHz gain residual template \citep{planck2016-l02}.}
    \label{fig:diff_30_k}
\end{figure}

\begin{figure}[t]
    \center
    \includegraphics[width=\linewidth]{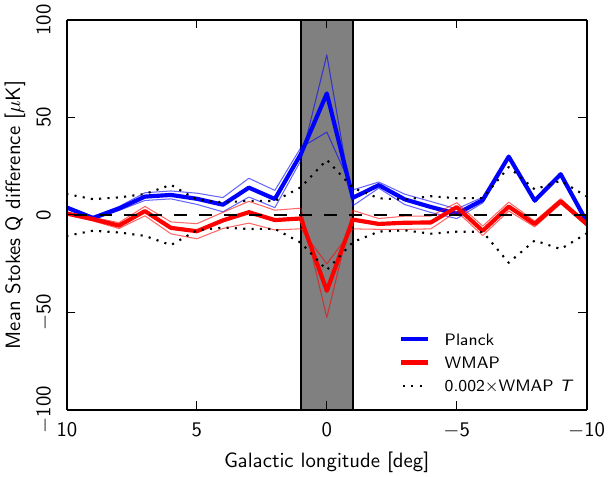}
    \caption{Latitude-averaged Stokes $Q$-band differences between
      QUIET\ 43\,GHz and \WMAP\ $Q$-band (red) and between
      QUIET\ 43\,GHz and \Planck\ 2015 44\,GHz (blue) over the
      QUIET\ Galactic center field, evaluated over a latitude band
      around the Galactic plane of $|b| \leq 1\pdeg5$; reproduced from
      \citet{ruud:2015}. The colored regions indicate the absolute
      QUIET\ calibration uncertainty of $\pm$6$\,$\%. The dotted
      lines show the latitude-band-averaged \WMAP\ Q-band temperature
      amplitude scaled by a factor of $\pm 0.002$, providing a rough
      template for a 0.2\,\% temperature-to-polarization leakage in \Planck. The
      gray region marks an area in longitude $\pm 1\deg$ around the
      Galactic center within which all results are dominated by
      uncertainties in the foreground spectral index.}
    \label{fig:quiet}
\end{figure}

\subsection{Breaking degeneracies through joint analysis of complementary datasets}

The fundamental motivation for the \BP\ and \cosmoglobe\ projects derives
directly from the experiences and insights gained within the \Planck\ project.
Towards the end of that project, it became clear that the main limiting factor
with respect to constraining large-scale CMB polarization comes neither from
instrumental systematics nor astrophysical foreground modeling as such but
rather from the interplay between the two\footnote{
  \href{https://www.cosmos.esa.int/web/planck/lessons-learned}
  {\texttt{https://www.cosmos.esa.int/web/planck/lessons-learned}}}. As
formulated in \citet{bp01}, \begin{quotation}
\ldots \emph{one cannot robustly characterize the astrophysical sky without
  knowing the properties of the instrument, and one cannot
  characterize the instrument without knowing the properties of the
  astrophysical sky}.
\end{quotation}

One demonstration of this ``chicken-and-egg'' problem is shown in
Fig.~\ref{fig:diff_30_k}, reproduced from \citet{bp07}, where the top panel
shows the Stokes $Q$ difference map between \Planck\ 2018 30\,GHz
\citep{planck2016-l02} and \WMAP\ $K$-band \citep{bennett2012}, after scaling
the latter by 0.495 to account for the spectral index of synchrotron emission.
Here one can see coherent large-scale patterns that massively dominate over the
random noise. The origin of these structures is well understood and can, to a
considerable extent, be described by the sum of transmission imbalance
uncertainties in \WMAP\ (middle row; \citealp{jarosik2007}) and gain
uncertainties in \Planck\ (bottom row; \citealp{planck2016-l02}). Both the
transmission imbalance and the gain estimation rely directly on knowledge about
the CMB sky, while estimating the CMB sky relies on knowledge about the
transmission imbalance and gain parameters. However, even though their physical
origins are well understood, they are still exceedingly difficult to mitigate
within each experiment individually, simply because the observation strategy of
each experiment leaves them nearly blind to these particular modes. One of the
solutions to this problem is to jointly analyze each experiment and use the
information in one experiment to break the degeneracies in the other.

Such complementary information need not only come from expensive satellite
missions but can also come from less expensive ground-based experiments. One
example of this is depicted in Fig.~\ref{fig:quiet}, reproduced from
\citet{ruud:2015}, which shows latitude-averaged polarization differences
between the 43\,GHz QUIET\ map and the corresponding 9-year \WMAP\ (red curve)
and \Planck\ 2015 (blue curve) maps. In this case, one can see an excess in
\Planck\ with respect to QUIET\  outside $|b|<1^{\circ}$, while QUIET\ and
\WMAP\ agree well. The most likely origin of this excess is bandpass-induced
temperature-to-polarization leakage \citep{ruud:2015,bp09} in the \Planck\ map
at the level of 0.2\,\% (dotted line), which is fully consistent with the
quoted systematic error of this channel of $<$1\,\% \citep{planck2014-a04}. To
reduce this systematic uncertainty below that achievable by \Planck\ alone,
additional external information is required, and the deep and highly
cross-linked Galactic plane measurements made by QUIET\ are precisely what is
needed for this.

These are only two relevant examples, and many more could be listed.
Nevertheless, such examples motivate our contention that in order to break
instrumental and astrophysical degeneracies, all free parameters should be
optimized jointly while simultaneously exploiting as many state-of-the-art
complementary datasets as possible. As such, the analysis problem should be
solved globally, both in a statistical and a research community sense. 

\subsection{The \BP\ data model and posterior distribution}
\label{sec:posterior}

As a first proof-of-concept of this global analysis approach, the \BP\
collaboration \citep{bp01} was formed with the explicit goal of re-analyzing
the \Planck\ LFI measurements. This data set represents a significant, but
manageable challenge in terms of data volume and systematics complexity. Also,
building on the experiences gained through the \Planck\ project, we chose to
adopt standard Bayesian parameter estimation techniques for our computer codes
because of their unique flexibility and fidelity in terms of systematic error
propagation. In particular, in the interest of saving costs and development
time, we chose the CMB Gibbs sampler called \commander\
\citep{eriksen:2004,eriksen2008,seljebotn:2019} as our starting point, which
formed a cornerstone in the \Planck\ data analysis
\citep{planck2016-l04,planck2016-l05,planck2016-l06}.

The most crucial component in any Bayesian analysis is a parametric model for
the data, which may typically take the following symbolic form,
\begin{equation}
\d = \s(\omega) + \n,
\end{equation}
where $\d$ denotes a given dataset, $\s(\omega)$ is a signal model
with free parameters $\omega$, and $\n$ is noise. The posterior
distribution, which quantifies the probability distribution of
$\omega$ as constrained by $\d$, is given by Bayes' theorem,
\begin{equation}
  P(\omega\mid \d) = \frac{P(\d\mid \omega)P(\omega)}{P(\d)} \propto
  \mathcal{L}(\omega)P(\omega),
  \label{eq:jointpost}
\end{equation}
where $\mathcal{L}(\omega)\equiv P(\d\mid \omega)$ is called the
likelihood, and $P(\omega)$ is called the prior; $P(\d)$ is a
normalization constant that is irrelevant for our purposes. The
likelihood quantifies the constraining power of the actual data, while
the prior summarizes our pre-existing knowledge regarding $\omega$
before the analysis.

For the \Planck\ LFI analysis that is presented in a series of
companion papers, we have adopted a parametric model that takes the
following form,
\begin{equation}
  \begin{split}
    d_{j,t} & = g_{j,t}P_{tp,j}\left[B^{\mathrm{symm}}_{pp',j}\sum_{c}
      M_{cj}(\beta_{p'}, \Delta_{bp}^{j})a^c_{p'}  + B^{\mathrm{asymm}}_{pp',j}\left(s^{\mathrm{orb}}_{j,t}  
      + s^{\mathrm{fsl}}_{j,t}\right)\right] \\
    & + s^{\mathrm{1\mathrm{Hz}}}_{j} + 
    n^{\mathrm{corr}}_{j,t} + n^{\mathrm{w}}_{j,t}.
  \end{split}
  \label{eq:todmodel}
\end{equation}
For the purposes of the current paper, the specific meaning of each symbol is
irrelevant, and we, therefore, refer the interested reader to Sect.~7 in
\citet{bp01} for complete details. Here it is sufficient to note that this
expression represents an explicit parametric model of both the instrument, as
quantified by $\{g, P, B, \Delta_{bp}, s^{\mathrm{fsl}}, n^{\mathrm{corr}}\}$,
and the astrophysical sky as expressed by the sum over components $c$, which
for \BP\ includes CMB, synchrotron, free-free, spinning and thermal dust
emission, and compact sources.

It is important to note that this model is far richer than what \Planck\ LFI is
able to constrain on its own. Simply by counting degrees of freedom alone, we
immediately note that the sky model has five free component amplitude
parameters per pixel, while the LFI data only provide three independent
frequencies. Consequently, the model is massively degenerate, and the LFI data
must be augmented with external data. This is done in the current \BP\ analysis
by including the \textit{WMAP} 33--61\,GHz \citep{bennett2012}, \Planck\ HFI
353 and 857\,GHz \citep{planck2016-l03}, and Haslam 408\,MHz \citep{haslam1982}
measurements in the form of pixelized frequency maps. The advantage of this is
that the model is now reasonably well constrained -- but a major disadvantage
is that these external pixelized sky maps may be associated with their own
systematic errors that may compromise the final results. To fully exploit the
strengths of each dataset in breaking degeneracies through joint analysis, one
should ultimately start from raw TOD for all involved observations, and
properly model all potential systematic effects. This is a main goal of the
\cosmoglobe\ effort.

\subsection{Low-level systematics, data longevity,
  and cost optimization}

Whenever the signal-to-noise ratio of a given dataset increases, new systematic
effects become important and must be modeled. This general observation also
holds true for the CMB community, which currently targets signals at the
nanokelvin level; even minuscule effects need to be accounted for at such low
signal levels. This directly increases the importance of external data, as no
planned experiment is able to measure all relevant effects internally at the
required precision. More typically, each experiment focuses on one piece of the
entire puzzle that it does particularly well for technological reasons and
relies on other experiments to provide information regarding other free model
parameters.

At the most basic level, the reason for this optimization is just a matter of
cost and complexity. In particular, modern CMB experiments cost so much that it
is unacceptable for funding agencies and taxpayers to repeatedly and needlessly
measure the same quantities. For example, the ground-based or sub-orbital
experiments typically cost at least about \$10\,M and involve 20--50 people,
while current and next-generation satellite experiments usually cost hundreds
of millions of dollars and involve hundreds of people.

To achieve new transformative results in the future within realistic budget
limits, these existing million-euro investments must be optimally leveraged and
re-used for all future experiments. For this to be possible, however, it is
also vital that the systematic error budgets of the old datasets are consistent
with the requirement of the new experiments. Unfortunately, this has
traditionally been a prohibitive challenge for one simple reason: Until now,
most CMB experiments have primarily published frequency maps or angular power
spectra---that are static by nature---as their main products. Once the
time-ordered data have been co-added into pixelized maps, it is no longer
possible to account for a wide range of low-level systematic uncertainties, but
only a very limited number of high-level uncertainties, such as white noise,
correlated noise on large angular scales \citep{bennett2012,bp10}, a single
absolute calibration factor \citep{planck2014-a10,bp07}, or symmetrized beam
uncertainties \citep{planck2016-l05}. This significantly limits the use of
legacy data for future analyses.

There are two noteworthy exceptions to this rule, though, namely \Planck\ and
\WMAP. Both published their full uncalibrated time-ordered data as part of
their legacy releases. Hence, the corresponding co-added frequency maps may, at
least in principle, be continuously improved as new information and
complementary datasets become available. However, it is also important to note
that neither \Planck\ nor \WMAP\ released \emph{the data analysis pipelines}
that were used to reduce the data.

That is problematic for at least two reasons. First, from a practical point of
view, the lack of functional analysis pipelines makes it very difficult and
time-consuming for external scientists to repeat and improve the original
analyses. Even more problematic, however, is the fact that most modern data
reduction pipelines typically employ a significant number of critical ancillary
datasets, for instance, ADC correction tables or far-sidelobe models. Since
these are only used during low-level processing, few external researchers ask
for them. As a result, they may easily be forgotten during the last stages of
the main collaboration work and sometimes even lost when the original
production computer systems are discontinued.

We argue in this paper that the optimal -- if not only -- way to ensure full
reproducibility and data longevity is to release a complete and functional data
processing pipeline together with the raw data, parameter files, high-level
products, and documentation. Furthermore, we also argue that such a complete
release should be required and supported in terms of dedicated funding for all
future CMB experiments by the respective agencies (ESA, NASA, JAXA, NSF, etc.).
This is clearly also in the funding agencies' own interests, as it guarantees
that their investments may be optimally leveraged in future work.

In addition to sharing data, it is also worth noting that sharing analysis
tools may lead to cost optimization of any given new experiment. Indeed,
establishing common analysis tools across the field will free up analysis
funding that can be better spent on understanding the instrument, exploring
ground-breaking theories, or deriving novel secondary science. An important
pioneering CMB-related example of this is \healpix\ \citep{gorski2005}, which
both defines a standard pixelization that facilities easy data sharing and
comparison across experiments, and provides a wide range of state-of-the-art
and user-friendly tools to operate on these data \citep[e.g.,][]{zonca2019},
all published under an Open Source license. In general, common software tools
are highly beneficial for the science community, funding agencies, and
taxpayers.

\subsection{Open Science: From \BP\ to \cosmoglobe}

An important goal for the \BP\ project was to develop and publicly release a
complete end-to-end analysis pipeline for one of the essential datasets in
contemporary CMB cosmology, namely the \Planck\ LFI data \citep{bp01}. The
motivation for this was two-fold. The first aim was to resolve a few notable
issues with the LFI data that remained unresolved at the end of the official
\Planck\ mission, in particular related to the global estimation of the
instrumental gain \citep{planck2016-l02,bp07}. However, this represents only a
first step in a much larger process, as embodied within the \cosmoglobe\
program, whose goal is to develop a general low-level analysis pipeline that
would be applicable to a much more comprehensive range of experiments --
legacy, current, and future -- and at the same time support \emph{joint}
analysis of these.

The full scope of this project is massive. For this work to succeed in the long
term, it must be firmly based on an Open Science foundation: While a small
group of dedicated people may be able to re-analyze one experiment (as \BP\ has
done for \Planck\ LFI), integrating a wide range of complementary experiments
without community contributions is unfeasible for several reasons. First, some
important datasets may be proprietary, and the original stakeholders must be
leading the analysis for legal reasons alone. Second, the systematic properties
of most datasets are quite complicated, and expert knowledge is usually
essential to formulate and generalize the data model. Third, the sheer amount
of work to be done effectively requires cost-sharing among all interested
parties, recognizing the currently constrained funding environment most
researchers experience daily.

In general, \cosmoglobe\ will be a hub for which analysis of both time-ordered
and map-domain data from different experiments can be integrated into a single
framework. A critical goal of \cosmoglobe\ is to support scientists working on
incorporating the data from their own experiments into the larger \cosmoglobe\
framework and thereby analyzing the data efficiently, robustly, and
economically. For the casual user who might be primarily interested in what the
microwave sky looks like at some specified frequency, \cosmoglobe\ will provide
a state-of-the-art and user-friendly sky model. By allowing scientists easy
access and configuration of their experiment in this framework, \cosmoglobe\
will become an integral tool for forecasting and planning for experiments.

With a functional codebase in hand, as demonstrated by the current LFI-based
data release, we believe that the time is now right for all interested parties
to get involved in this work and extend the framework according to their own
needs. We anticipate that such contributions will most typically take one of
two forms. The first is stand-alone projects, in which the external user simply
downloads the software and data and performs some analysis without input from
the greater community. In this case, the only formal obligation for the user is
to publish all derived codes under an equally permissive software license as
the one used for \BP\ (which in practice means a GNU General Public License
(GPL)) and to acknowledge previous work through appropriate referencing.

The second mode of operation is active participation in the \cosmoglobe\
framework. In this case, an external user or project may request direct expert
\cosmoglobe\ support, for instance, in the form of software development and
data analysis assistance. That is then likely to increase the chance of success
significantly. In exchange for the support, the external user or project must
commit to publicly releasing the underlying data after some proprietary period,
and all directly contributing \cosmoglobe\ collaborators must be offered
co-authorship, in accordance with standard scientific practices. Required
details can be specified in a Memorandum of Understanding (MoU) between the
external party and the relevant \cosmoglobe\ participants before the work
commences. \cosmoglobe\ is intended to be a platform for initiating and
supporting mutually beneficial collaborations.

\section{Reproducibility survey, tools, and licensing}
\label{sec:reproducibility}

We now turn our attention to the practical aspects of how to build an Open
Science-based foundation for this work and examine some of the latest
developments on reproducibility in science in general. Next, we identify
several tools and services that are available online and aim to provide
solutions for reproducible science. Finally, we review the most popular
licenses used for Open Source development. These issues cover a variety of
topics that constitute the current state of the art in reproducibility and Open
Science as of 2018--19.

\subsection{Open Science development tools per 2018}

We collected information regarding available tools that might be useful to
strengthen the reproducibility aspects of the project. We found this exercise
quite informative and helpful, and we highly recommend future collaborations to
conduct similar meta-studies \emph{before} starting the data analysis work, as
it is easy to get swamped with scientific problem solving once the main effort
begins. We also note that the field moves very quickly, and the
state-of-the-art is likely to be quite different only after a few years.

\subsubsection{Workflow definition tools versus integrated software}

We first considered the use of so-called ``workflow definition tools'' to
organize the primary data model and Gibbs sampler discussed in
Sect.~\ref{sec:posterior}. Such workflow managers help scientists define and
execute a specific set of tasks, implemented by executing local (or sometimes
remote) code, scripts, and other sub-workflows. Each component is only
responsible for a small fragment of functionality. Therefore many pieces are
working together in a pipeline to achieve the ultimate goal of the workflow,
performing a useful task.

There have already been attempts at implementing and utilizing such tools in
the CMB community. The most well-known example is the \texttt{ProC} workflow
manager\footnote{\href{http://planck.mpa-garching.mpg.de/ProC/}
{\texttt{http://planck.mpa-garching.mpg.de/ProC/}}} developed by the Max Planck
Institute of Astrophysics for the \Planck\ mission. Two popular Open Source and
general-purpose tools are
\texttt{Taverna}\footnote{\url{https://taverna.incubator.apache.org}} and
\texttt{The Kepler Project}\footnote{\href{https://kepler-project.org/}
{\texttt{https://kepler-project.org/}}}.

The main advantage and attraction of such workflow managers is their ability to
construct complex, flexible, and reproducible workflows based on well-defined
components. At the same time, what makes these managers work is very strict
interfaces between the different components. Unfortunately, this strictness
adds significant additional burdens on the code developers, both in terms of a
steep learning curve to be able to add new features and in terms of restricted
flexibility to implement new solutions to unexpected problems; it is difficult
to make substantial changes without breaking compatibility with already
existing components. A second significant challenge is efficient memory
management. Suppose the various pipeline components are written in different
programming languages. In this case, one must either resort to data sharing
through slow disks or spend great effort on highly non-trivial in-memory
communication.

In general, our experience is that general-purpose workflow managers tend to be
more practical for well-established and relatively quick routine tasks than for
cutting-edge research that relies on high-performance computing. The most
critical priority for our Bayesian analysis framework is computational speed,
as a factor of six in runtime can make the difference between a two-month
runtime (which is painful but doable) and one year (which is prohibitive).
Optimal memory and disk management are, therefore, the key. The second most
important priority is code flexibility, allowing developers to introduce
changes needed to achieve their goals freely.

After careful consideration, we decided to drop the use of workflow managers,
as the official \Planck\ Data Processing Centers (DPCs) did, to maintain
optimal coding agility and flexibility. However, in contrast to the \Planck\
DPCs, we instead opted for developing the entire analysis pipeline within one
single computer program called \commanderthree, to ensure optimal memory
management and computational speed \citep{bp03}. Two important additional
advantages of implementing the entire pipeline within a single code are that
the whole collaboration naturally develops a common ``language'' and knowledge
base that are helpful to discuss issues more efficiently, and it also minimizes
duplication of effort.

\subsubsection{Online development services}

Another potentially useful class of tools is the so-called ``online development
services''. These offer the possibility to perform all development work to be
done online through the use of general-purpose web applications. Some of the
major players in this area that we evaluated were \texttt{Open Science
Framework}\footnote{\url{https://osf.io/}},
\texttt{Codeocean}\footnote{\url{https://codeocean.com/}}, and
\texttt{Zenodo}\footnote{\url{https://zenodo.org/}}.

One of the major advantages of these services is that, by definition, all work
is performed online. This facilitates very easy dissemination since results may
be published in an Open Science manner literally in real-time. However, our
evaluation is that they are also associated with three main disadvantages,
mirroring the issues discussed in the previous section. First and foremost,
online services typically impose a specific and inflexible work style that may
not suit everybody in a large collaboration. Second, they have a significant
learning curve that may be off-putting to many scientists with busy schedules.
Third, depending on the plans offered by each provider, authors can easily run
out of hosting space or online computational time, requiring them to update
their accounts. While this may make financial sense from the side of the
hosting companies, we consider this to be a big disadvantage for authors, just
for reproducibility purposes alone. A solution like this might make sense for
small projects, but the cost can become prohibitive for larger and heavier
collaborations.

At their current stage of development, we, therefore, also decided to avoid the
use of integrated online development services and instead leave each
collaborator to choose their own development environment individually. We also
note that most scientists are, by nature, quite independent-minded and do not
necessarily respond well to being imposed on a specific development
environment. However, if the available tools offered more obvious advantages,
the situation might be different, and we definitely recommend future
collaborations to perform a similar survey.

\subsubsection{Software repositories}

One class of software development tools that is critical for a large-scale Open
Source effort is efficient \textit{revision control systems} (RCSs). This
allows users to collaborate on the same computer program or scientific paper in
real-time with a minimum of synchronization problems and is a cornerstone of
modern software development. As noted above, 56\,\% of the user survey
respondents already use at least one such system, with \texttt{Git} being the
most popular.

At the beginning of the program, we quickly settled on \texttt{Git} as our main
RCS, primarily because it was most widespread in our group, but also because we
find that it handles merges and conflicts better than most competitors. The
main question was then which (if any) common repository we should use. Three
particularly well-known providers are
\texttt{Bitbucket}\footnote{\url{https://bitbucket.org}},
\texttt{GitHub}\footnote{\url{https://github.com}}, and
\texttt{GitLab}\footnote{\url{https://about.gitlab.com/}}.

One advantage of \texttt{GitLab} and \texttt{Bitbucket} is that they offer free
private repositories in addition to public ones. This option might make them a
better candidate for users who want to start their project as a private
repository but switch to a fully public repository later on, closer to
publication time. In addition, all three offer a full suite of online
development tools, including bug/issues management, wiki pages, file hosting
capabilities, and API access to hosted files.

Initially, we adopted \texttt{GitLab} as our main provider, primarily because
it allows code to be run remotely on their web hosts. In addition, we
considered that it might be helpful for small tasks, such as implementing
online tools and calculators or automatically compiling paper drafts after each
submission. However, it is important to note that this feature is only free for
limited usage. Therefore, we have concluded it was not as useful as initially
anticipated. Consequently, halfway through the project, we have switched to
\texttt{GitHub} for our central software repository, simply because most people
in our community already have accounts there and to avoid overhead by
maintaining two separate accounts for most users.

\subsubsection{Open Source licenses}

When working in an Open Science setting, it is vital to protect the investments
and interests of the various contributors and users. A critical aspect of this
is licensing. Today, many Open Source licenses are in active use, and an
important task for projects like \BP\ and \cosmoglobe\ is to choose the
appropriate one for the work at hand. In this section, we provide a brief
overview of licenses in the most common use today and discuss which one was
chosen for our project, given the basic requirements that (1) our software
should be Open Source; (2) all derivatives of this work should remain Open
Source; and (3) our license should not contradict the licenses of any
dependencies (\texttt{HEALPix}\footnote{\url{http://healpix.sourceforge.net} or
\url{https://healpix.sourceforge.io}},
\texttt{FFTW3}\footnote{\url{https://www.fftw.org}}, etc.).

Although the term \textit{Open Source software} may be intuitively understood
to be freely distributable, modifiable, and shareable code written by a single
or a group of programmers, it is, in reality, more complex than it may seem at
first glance. Generally speaking, when discussing software licenses, one needs
to distinguish between several different aspects and concepts. The first aspect
concerns basic distribution. On the most restricted side, \emph{proprietary
software} is considered private property, and users are not allowed to share,
study, change, or reverse engineer the provided software. A variation of this
is called \emph{freeware}; in this case, the original software developer
retains all rights, and the only difference is that end-users do not need to
pay for the basic usage of a given program. \textit{Source available software}
is software that allows users to view the source code but does not necessarily
give the right to distribute, modify and/or install it on their machines. One
example of such a license is the \textit{Commons Clause
License}\footnote{\url{https://commonsclause.com/}}, which prohibits users from
selling the software. Because of such restrictions, source available licenses
are generally not considered to be Open Source. Next, \textit{Public Domain
software} or ``unlicensed'' software is software that waives all the rights of
copyright, trademark, or patent. Such software belongs to the ``public'' that
uses it, and it can be freely distributed, modified, and/or sold without
attribution to anyone. Examples of such licenses are \textit{Creative Commons}
(CC0) and \textit{Unlicense}. During the course of history, experts have been
arguing whether this type of license should be considered Open Source or
not. Indeed, some argue that the fundamental rights of free and Open Source software are not guaranteed\footnote{For more details, see, e.g., the “Public Domain Is Not Open Source” article: \url{https://opensource.org/node/878}} since the uncopyrighted software may be a subject to restrictions depending, e.g., on the laws of the particular country. Some are going as far as to claim that Public Domain is not at all an appropriate license\footnote{Lawyer Lorence Rosen has written the essay titled ``Why the public domain isn't a license'' --- faced with strong opposition, he has accepted that CC0 can be considered Open Source.} for computer software. Because of these disagreements, we also do not consider this license to be a ``proper'' Open Source license in the present work.

Moving on to what is considered proper Open Source software, \textit{free
software} (FS), as defined by the \textit{Free Software
Foundation}\footnote{\url{https://www.fsf.org/}} (FSF), is ``software that
gives its users the freedom to run, copy, distribute, study, change and improve
the software''.  According to a formal definition, the software is not
considered free if it does not respect the following four essential
freedoms:\footnote{\href{https://www.gnu.org/philosophy/free-sw.html}
{\texttt{https://www.gnu.org/philosophy/free-sw.html}}}

\textbfit{Freedom 0:} The freedom to run the program for any
purpose.

\textbfit{Freedom 1:} The freedom to study how the program works,
and change it so it does your computing as you wish. Access to the
source code is a precondition for this.

\textbfit{Freedom 2:} The freedom
to redistribute copies so you can help others.

\textbfit{Freedom 3:} The
freedom to distribute copies of your modified versions to others. By
doing this you can give the whole community a chance to benefit from
your changes. Access to the source code is a precondition for this.

Finally, \textit{Open Source licenses} are defined by the \textit{Open Source
Initiative}\footnote{\url{https://opensource.org/licenses}} (OSI) and include
licenses that ``allow the software to be freely used, modified, and shared'',
and comply with ten distinctive
criteria\footnote{\url{https://opensource.org/osd}} that concern (1) free
redistribution; (2) source code; (3) derived works; (4) integrity of the
author's source code; (5) no discrimination against persons or groups; (6) no
discrimination against fields of endeavor; (7) distribution of license; (8) the
license must not be specific to a product; (9) the license must not restrict
other software; and (10) the license must be technology-neutral.

\begin{figure*}[ht]
    \center
    \includegraphics[width=\linewidth]{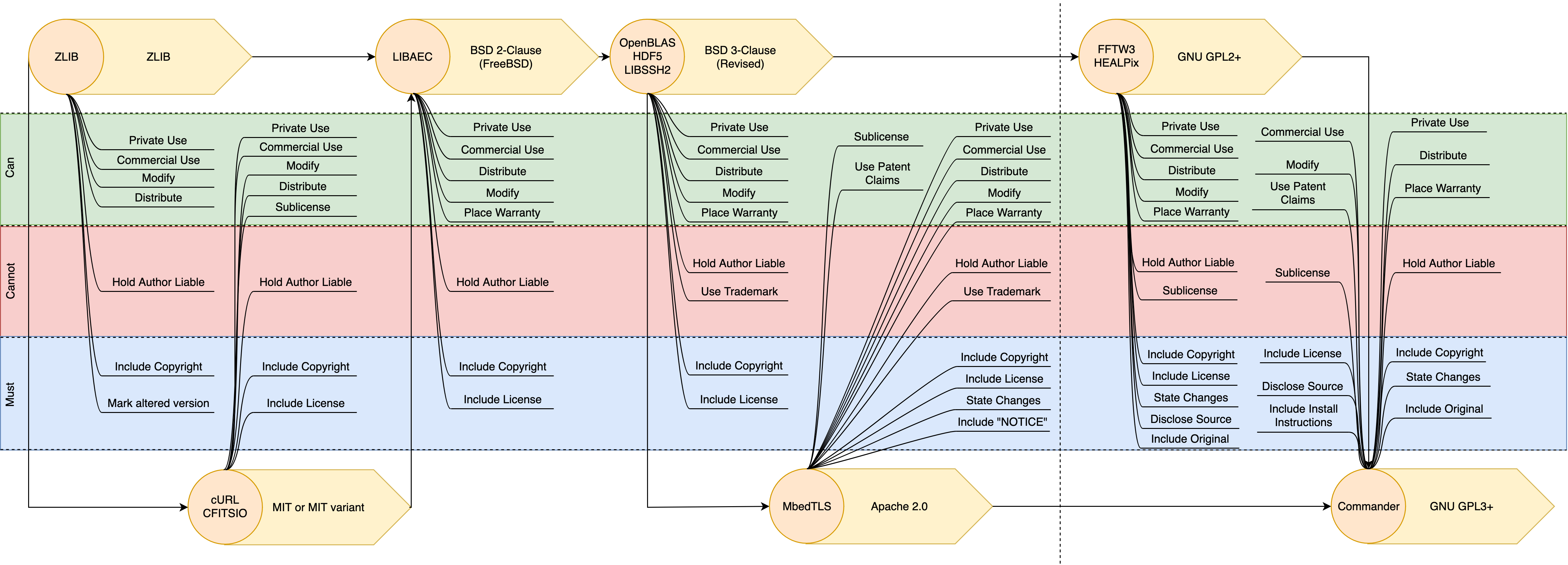}
    \caption{Licensing diagram
      for \commander\ and its dependencies,
      ordered from left to right according to increasingly specific
      licenses. Circles indicate libraries and codes, while rectangles
      show the licenses under which the particular library was
      issued. The dashed line divides the so-called permissive (i.e.,
      licenses that can be combined and used together with proprietary
      software) and restrictive licenses (i.e., licenses that enforce
      the source code to stay open for general public). This diagram 
      was derived from the David A. Wheeler's 
      \href{https://dwheeler.com/essays/floss-license-slide.html}{FLOSS License Slide}
      and the information provided on the \url{https://tldrlegal.com/}.} 
    \label{fig:licensing_diagram}
\end{figure*}

Both FS and OSI are considered to be Open Source with only subtle
differences.\footnote{These are more philosophical in nature. As
  Richard Stallman, the founder of GNU Project and FSF, stated: \textit{``The
  term `open source' software is used by some people to mean more or
  less the same category as free software\ldots The differences in
  extension of the category are small: nearly all free software is
  open source, and nearly all open source software is free.''}} While FS
is focused on the user's rights to use, modify and share the program,
OSI is focused on the source code being open with unrestricted
community driven
development.\footnote{\url{https://opensource.org/about}} Since the main
goal of the \cosmoglobe\ project is to build a community around a
common source code, this strongly suggests that the OSI definition is
most suitable for our purposes. However, looking at OSI's list of
the most popular and widely used licenses,\footnote{For the full list
  of open-source licenses, please visit:
  \newline
  \url{https://opensource.org/licenses/category}}
which includes GNU variants, we are going to use both definitions
interchangeably. 

In order to select one of these OSI licenses, it is important to recognize that
the \commander\ source code does not exist in a vacuum but rather depends
directly and indirectly on a variety of different libraries, as visualized in
Fig.~\ref{fig:licensing_diagram}. When one chooses the correct license for the
project, the list of dependencies must be considered. Then, typically, the most
specific one defines what is allowed for the new software. However, this is not
always the case. For example, \texttt{cURL}\footnote{
  \url{https://github.com/curl/curl/blob/master/COPYING}} is based on the
modified MIT license, but it may or may not be compiled with the support of
\texttt{MbedTLS}\footnote{
  \url{https://github.com/ARMmbed/mbedtls/blob/development/LICENSE}}
and \texttt{LibSSH2}\footnote{
  \url{https://github.com/libssh2/libssh2/blob/master/COPYING}} that
rely on more complicated licenses. The same applies to, e.g.,
\texttt{CFITSIO}\footnote{\url{https://heasarc.gsfc.nasa.gov/docs/software/fitsio/c/f_user/node9.html}}
which requires
\texttt{ZLIB}\footnote{\url{https://www.zlib.net/zlib_license.html}} starting
from version 4.0.0. A third important example is \commander\ itself, which may
be compiled with Intel Parallel Studio and Intel Math Kernel Library
(\texttt{MKL}), which are issued under one of Intel's proprietary licenses. We
do not ship or install this library together with \commander\ in any way, and,
thus, such a decision relies solely on the user's judgment.

Our understanding is that as long as one does not explicitly modify the source
code of a library issued under a permissive license, the licenses can be used
almost interchangeably. Generally, this does not apply to the restrictive ones.
However, the GPL licenses are internally compatible as long as it includes a
special line\footnote{ \textit{``(...) either version 2 of the License, or (at
your option) any later version. (...)''}}  that explicitly states that the
later version of the license can be used. In practice, this means that if a
piece of software is issued under GPLv2 and not GPLv2+, then we cannot use it.  

All in all, the license adopted for \commander\ and \cosmoglobe\ cannot be more
restrictive than the most permissive of these, which is set by the GPL2+
license adopted for \texttt{HEALPix}. In principle, the same applies to
\texttt{FFTW3}, but if needed, this library could have been replaced with other
FFT implementations. In contrast, \texttt{HEALPix} is in practice irreplaceable
and therefore determines the license also for the current work. Consequently,
our final choice is the GNU General Public License v3, as dictated by the
diagram. Of course, for us, this is not only a matter of formality but also of
preference; we \emph{want} this software to be and remain Open Source to
protect the interests of everybody involved. Therefore, it is crucial for all
future participants and developers of the \cosmoglobe\ framework to familiarize
themselves with this license and determine whether this is acceptable for one's
work and compliant with potential collaboration policies.

\section{Compilation support}
\label{sec:compilation_support}

The \commander\ software is primarily intended to be run on
\texttt{Linux}-based \textit{High-Performance Computing} (HPC) clusters with
basic \texttt{Fortran} and \texttt{MPI} compilers and libraries available and
typically tens to thousands of computing cores. Beyond that, it does not impose
any specific constraints on the computing platform, neither in the form of
processor architecture (AMD, ARM, Cray, Intel, etc.)  nor compilers (GNU,
Intel, PGI, etc.). At the same time, the code does depend on many external
libraries, including \texttt{HEALPix}, \texttt{FFTW3},
\texttt{CAMB},\footnote{\url{https://camb.info/}} and
\texttt{HDF5};\footnote{\url{https://www.hdfgroup.org}} for full specification,
see the online \commander\
documentation.\footnote{\url{https://cosmoglobe.github.io/Commander/}} The
combination of many dependencies and a rich computing platform heterogeneity
can, in general, represent a significant challenge and workload in terms of
compilation and can be seriously off-putting to many users. Therefore, it is
critical to make this process as simple and user-friendly as possible, and
automated build systems play a key role in that work.

Build systems come in various flavors and combinations, which makes it
non-trivial to choose one among many. As a result, there is no single ``best''
build system, but each has its advantages and disadvantages. Therefore, when
selecting one specific system for \commander, we have considered seven
different aspects listed in order of importance.

\textbfit{1) Free and Open-source}: \commander\ will not be a genuinely Open
Source tool if the build system which installs it is itself not free and Open
Source.

\textbfit{2) Cross-platform}: Although Linux was (and still is) the predominant
operating system for modern HPC, there is currently a trend toward a more
diverse landscape.\footnote{According to
\href{https://www.top500.org/statistics/list/}{top500.org}, the share of
non-Linux systems running the world's top HPC systems was 2.4\,\% in November
2010, 3.4\,\% in November 2014, 7\,\% in November 2019, and 16\,\% in November
2021.} In addition, Windows and Macintosh software dominate the PC and laptop
market, and these systems are growing rapidly in power and can be used today
for productive analysis. Hence, we require that the build system must allow us
to compile and run \commander\ on any major operating system, including Linux,
Windows, and MacOS.

\textbfit{3) Automation}: Many astrophysics departments operate today their own
computing cluster. However, experience shows that there is often limited
professional support for the installation, tuning, and maintenance of software
and libraries. This work is often left to the scientists who want to run the
code; Ph.D. students, postdocs, and professors. The installation procedure must
therefore be both transparent and easy to use. Furthermore, it should be
automated, i.e., the build system should be able to check for the existence of
specific \commander\ dependency, and, if it doesn't find it on the host system,
it should download, verify, compile, and install all missing dependencies,
compile \commander\ itself, and link them all together.

\textbfit{4) Long-term support:} While Open Source projects have many
advantages in current computing, they also tend to have one fundamental flaw:
People tend to abandon them in favor of ``newer'' and ``better'' codes. This
poses an obvious threat to a large and long-term project such as \cosmoglobe,
which will require significant community-wide investments over many years. For
this reason, the adopted build system should be mature and supported by a
dedicated community. Proven stability is more important than cutting-edge.

\textbfit{5) Multi-language support:} The bulk of the \commander\ codebase
consists of Fortran code, but there are many \texttt{Python} scripts and
libraries written in the course of the development. In addition, there are also
several \texttt{C++}-based modules that need to be compiled together with
\commander, and other languages may become useful in the future. Therefore, the
tool should have support for multiple languages used simultaneously in one
project.

\textbfit{6) Minimal dependencies:} The build system should have as few
dependencies as possible, and these should ideally either be already present in
the system or easy to install from the source. Preferably, it should be a
single binary or a piece of software.

\textbfit{7) IDE Integration:} While \textit{Integrated Development Enviroment}
(IDE) support is not a strict requirement, it is certainly a nice bonus
feature. In the present day and age, people are using a huge variety of IDEs
that can perform syntax highlighting and code checking—having the same features
available for the build system gives the programmer an advantage in terms of
the code development speed.

With these points in mind, we now provide a survey of possible useful
compilation support tools in current use and then describe the implementation
chosen for \commander.

\subsection{Survey of automatic build systems}
\label{sec:build_systems_overview}

\subsubsection{Low-level build systems -- \texttt{Make}}
\texttt{Make} is a ``low-level'' build automation tool which uses special
instruction files -- so-called \texttt{Makefiles} -- to build and install
software from the source code. It has a variety of implementations, with
perhaps, the most widespread one being GNU \texttt{Make} which is shipped
together with most \texttt{Linux} and \texttt{Unix} distributions. In fact,
\commander1 and \commander2 were solely built using \texttt{Make}. Its
advantages are numerous: It is a standard Linux and Unix tool; it is
widespread, and the majority of the scientific and open source community knows
about it; it will not be deprecated in the foreseeable future since it has a
solid community of maintainers; it has a support for a variety of languages such
as \texttt{C}, \texttt{C++}, \texttt{Fortran}, \texttt{Java}, \texttt{Python},
etc.; it allows nested project structures; and, starting from version 3.0, it
allows compilation using multiple processes.

However, it also has a few notable disadvantages. First, \texttt{Make} has a
relatively obscure syntax and associated steep learning curve. Second, for
large projects, such as \commander, the \texttt{Makefiles} tend to become very
long and complex and increasingly hard to maintain. Finally, code compilation
requires specific instructions for each particular compiler, OS, and
architecture, making it a poor solution for cross-platform development. Despite
these shortcomings, \texttt{Make} remains the most widespread build system in
use today, and it has a firmly established user community. Taking into account
both its flexibility and maturity, we have chosen \texttt{Make} as the primary
low-level compilation system for \commander.

\subsubsection{High-level build system -- \texttt{CMake}}

Developed since 1999 under a BSD-3-Clause license, \texttt{CMake} is a ``meta''
or ``high-level'' build system that is used in conjunction with some other
``low-level'' build environments, for instance \texttt{Make}, \texttt{Ninja},
or \texttt{Microsoft Visual Studio}. \texttt{CMake} may be used to build a
software project in a two-step process: First, \texttt{CMake} reads in a series
of configuration files written in \texttt{CMake}'s own scripting language,
called \texttt{CMakeLists.txt}, and use these to automatically produce a
complete low-level build system configuration (e.g., \texttt{Makefiles}).
Second, these files are used by the low-level native generator (e.g.,
\texttt{Make}) to actually compile and install the project.

\texttt{CMake} allows for flexible project structures. For instance, nested
directory hierarchies and/or complex library dependencies do not pose a
problem, as it can locate a variety of files and executables on the host
system. Once such dependencies have been identified, the location data are
stored inside a special file called \texttt{CMakeCache.txt} that can be
manually tuned before the actual build. In addition, it has the functionality
to download, verify, unpack and compile archives of missing libraries that
utilize non-\texttt{CMake} build systems. Furthermore, cross-compilation is
straightforward since \texttt{CMake} has extensive OS, language, and compiler
support.\footnote{\url{https://cmake.org/documentation/}} Other important
features include, but are not limited to, support for mathematical expressions;
string, list, and file manipulation; conditions, loops, functions, and macros;
and shell scripting.

Today, \texttt{CMake} is the de facto standard tool for \texttt{C++} project
development,\footnote{According to a 2019 survey by
\href{https://www.jetbrains.com/lp/devecosystem-2019/cpp/} {Jet Brains}, 42\,\%
of \texttt{C++} projects use \texttt{CMake}; 33\,\% use \texttt{Makefiles};
9\,\% use \texttt{Qmake}; 8\,\% use \texttt{Autotools}; and only 1\,\% use
\texttt{SCons}. In addition, according to the
\href{https://isocpp.org/blog/2021/04/results-summary-2021-annual-cpp-developer-survey-lite}{2021
Annual C++ Development Survey} roughly 80\,\% of all projects use
\texttt{CMake} as one of their build systems.} and a variety of Open Source
projects use \texttt{CMake} as its build system, including several \commander\
dependencies such as \texttt{CFITSIO}, \texttt{FFTW3}, and \texttt{HDF5}.
Lastly, \texttt{CMake} has good support for IDE integration.

However, despite its many strengths, \texttt{CMake} is not perfect. The syntax
has a pretty steep learning curve, and the source code may quickly become
cumbersome and difficult to read. Also, there does not seem to be a universal
approach, or even strict guidelines, for how to structure \texttt{CMake} code
for large and complicated projects. Finally, the documentation is extensive but
non-trivial to navigate and read for newcomers.

Based on its prevalence, mature community, rich feature set, and the fact that
many \commander\ dependencies also use \texttt{CMake}, we have chosen this as
our primary high-level build system, with \texttt{Make} as the corresponding
low-level system. Specific details regarding the \commander\ \texttt{CMake}
configuration is described in Sect.~\ref{sec:cmake}.

\subsubsection{Alternative build systems}

This section provides an overview of other competing systems that were explored
during the initial phases of the project but ultimately not selected. However,
several of these may be attractive candidates for future astrophysics and
cosmology Open Source projects.

\textbfit{\texttt{Ninja}}\footnote{\href{https://ninja-build.org}{\texttt{https://ninja-build.org}}}
is a low-level build system, similar to \texttt{Make}, specifically designed
for speed. Like \texttt{Make}, it supports a variety of languages, platforms,
compilers, and operating systems and, in fact, is meant to eventually replace
\texttt{Make}. The main reason for not choosing \texttt{Ninja} over
\texttt{Make} is simply the fact that it was designed to be used in combination
with high-level build systems and not on its own. In addition, it is still in
active development, and this carries a risk of higher -- and unnecessary --
development overheads for our purposes. This choice is likely to be revisited
in the future when \texttt{Ninja} has proven itself further in terms of
stability and user base.

\textbfit{\texttt{QMake}} or \texttt{makemake} is a build system created by the
Qt Company which automates the creation of \texttt{Makefiles}, similar to
\texttt{CMake}. It supports multiple platforms and can produce
\texttt{Makefiles} tuned for specific operating systems. Although mostly used
for \texttt{C++} projects, it can incorporate custom compilers (e.g.,
\texttt{gfortran}), and this allows it to work with \texttt{Fortran} source
files. However, it lacks native support for non-\texttt{C++} languages, a
natural alternative to \texttt{Make} build tools, and the ability to
incorporate third-party libraries directly into the build.

\textbfit{\texttt{XMake}} is a lightweight build system that supports multiple
languages, tool-chains, and platforms, and it can compile projects both
directly and produce configuration files for low-level build systems such as
\texttt{Make} or \texttt{Ninja}. However, native \texttt{Fortran} support was
added as recently as July 2020 (in version 2.3.6), well after a system had to
be chosen for the current \commander\ development. It is fast and has many IDE
plug-ins.

\textbfit{GNU \texttt{Autotools}} is a GNU build system composed of several
utility programs that are designed to make source code portable to many
\texttt{Unix}-like Operating Systems. It is widely used by many free and Open
Source projects, including several astrophysical and cosmology ones. In
general, \texttt{Autotools} generate the distribution archive used to build
programs. Once users obtain this package, they need to unpack it and run three
simple and well-known commands -- \texttt{configure}, \texttt{make} and
\texttt{make install} -- to compile and install the code using the facilities
provided by their host systems. Such an approach, in theory, eliminates the
need to install \texttt{Autotools} entirely, but, in practice, \texttt{Linux}
distributions still have it with (sometimes) multiple versions installed, which
adds to the confusion. Furthermore, a major drawback of \texttt{Autotools} is
its complexity; it requires much experience and time to develop robust and
user-friendly configuration files. For many future projects, we consider
\texttt{CMake} to be a more accessible and user-friendly solution.

\textbfit{\texttt{Scons}} does not implement a new special-purpose and
domain-specific language but rather utilizes specific \texttt{Python} scripts
to build the projects. Thus, the only requirement is to have \texttt{Python}
installed on the system, and this makes \texttt{Scons} cross-platform by
default; any system that runs \texttt{Python} can install \texttt{Scons} using
\texttt{Python}'s standard installation frameworks, \texttt{pip} or
\texttt{conda}. MIT license, good multi-language support, reliance only on
\texttt{Python}, and a rich feature set make it a tool worthy of exploring for
new projects. A major shortcoming is that there does not appear to be a simple
way of integrating other projects or dependencies into the build.

\textbfit{\texttt{Waf}} is another build system solely based on
\texttt{Python}. Since \texttt{SCons} inspired \texttt{Waf}, both tools have
many similar features: They rely exclusively on \texttt{Python}; are
cross-platform; can automatically scan for project dependencies; and have
support for multiple languages, including \texttt{C}, \texttt{C++} and
\texttt{Fortran}. However, while \texttt{SCons} is older and therefore is used
in more projects and has better documentation, \texttt{Waf} seems to be much
faster and provides more user-friendly console output that makes it easier to
debug. Additionally, it does not require a separate installation since the tool
is designed to be shipped as part of the main project source code.
Unfortunately, similar to \texttt{SCons}, there is no easy way to incorporate
other projects into the build.

\textbfit{\texttt{Meson}} is the last \texttt{Python}-based tool considered
here and is the closest \texttt{CMake} competitor we have found so far. It is
similar to \texttt{CMake} in many (if not all) aspects, and both are
meta-languages that compile the source code in a two-step process. However,
while \texttt{CMake} uses \texttt{Make} by default, \texttt{Meson} uses
\texttt{Ninja} instead; this makes \texttt{Meson} even faster in some cases. It
is also an Open Source, cross-platform tool that supports multiple languages,
including \texttt{Fortran}. In addition, nested hierarchies are not a problem,
as well as the incorporation of other projects (both \texttt{Autotools}- and
\texttt{CMake}-based) into the build.

\textbfit{Fortran Package Manager} or simply \texttt{fpm} is a relatively
new\footnote{ Alpha version was released on \texttt{GitHub} in November 25,
2020.} initiative by the Fortran-Lang
foundation\footnote{\href{https://fortran-lang.github.io/fpm/}{\texttt{https://fortran-lang.github.io/fpm/}}.}
inspired by \texttt{Cargo}, the package manager for the \texttt{Rust}
programming language. It is both a build system and a package manager that can
build libraries and applications. In addition, it has native support for unit
testing and can include other dependencies (e.g., from \texttt{Git}) into the
project. While looking very promising, we considered that it was not ready for
large-scale production at the time when the current project started. Still,
this option should be revisited in the future.

\subsection{\texttt{CMake}-based compilation}
\label{sec:cmake}

As discussed above, we have adopted \texttt{CMake} and \texttt{Make}
as our high- and low-level build systems. In this section, we provide
an overview of the \commander-specific \texttt{CMake} configuration
and compilation procedure.

\begin{figure}[t]
\tikzstyle{every node}=[draw=black,thick,anchor=west]
\tikzstyle{selected}=[draw=red,fill=red!30]
\tikzstyle{optional}=[dashed,fill=gray!50]
\begin{center}
      \begin{tikzpicture}[
                  grow via three points={one child at (0.5, -0.7) and
                              two children at (0.5, -0.7) and (0.5,-1.4)},
                  edge from parent path={[->](\tikzparentnode.south) |- (\tikzchildnode.west)}]
            \node {\texttt{/root}}
            child { node {\texttt{/cmake}}
                        child { node {\texttt{/compilers}} }
                        child { node {\texttt{/deprecated}} }
                        child { node {\texttt{/modules}} }
                        child { node {\texttt{/projects}} }
                        child { node {\texttt{/third\_party}} }
                        child { node {\texttt{libraries.cmake}} }
                        child { node {\texttt{main.cmake}} }
                        child { node {\texttt{sources.cmake}} }
                        child { node {\texttt{summary.cmake}} }
                        child { node {\texttt{toolchains.cmake}} }
                        child { node {\texttt{variables.cmake}} }
                  }
            child [missing] {}
            child [missing] {}
            child [missing] {}
            child [missing] {}
            child [missing] {}
            child [missing] {}
            child [missing] {}
            child [missing] {}
            child [missing] {}
            child [missing] {}
            child [missing] {}
            child { node {\texttt{/commander3}}
                        child { node {\texttt{src}} }
                        child { node {...} }
                  }
            child [missing] {}
            child [missing] {}
            child { node {\texttt{/docs}}}
            child { node {\texttt{/logo}}}
            child { node {\texttt{.gitignore}}}
            child { node {\texttt{CMakeLists.txt}}}
            child { node {\texttt{Makefile}}}
            child { node {\texttt{README.md}}};
      \end{tikzpicture}
\end{center}
    \caption{Overview of \commander\ source code directory structure. 
    The \texttt{CMake}-scripts are inside the \texttt{cmake} directory. Here, 
    \texttt{compilers} contains compiler configurations; 
    \texttt{deprecated} contains deprecated code for convenience; 
    \texttt{modules} contains custom written \texttt{Find<Name>.cmake} modules; 
    \texttt{projects} contains configuration for all library subprojects; 
    \texttt{third\_party} contains \texttt{CMake} modules taken from other projects; 
    \texttt{libraries.cmake} handles general subprojects' configurations; 
    \texttt{sources.cmake} contains urls and various hashes for the subprojects; 
    \texttt{summary.cmake} module is for debug output of the final \texttt{CMake} configuration 
    on the screen; 
    \texttt{toolchains.cmake} module handles general compiler configuration; 
    \texttt{variables.cmake} has custom-defined variables available for to build \commander\
    as a single project; 
    \texttt{main.cmake} is there to mimic the logic of \texttt{C++}, \texttt{Fortran}, 
    \texttt{Python} etc. codes, and serves as an analogy to the ``entry point'' of the program.}
    \label{fig:source}
\end{figure}

\subsubsection{\commander-specific \texttt{CMake}-code organization}

Usually, when one is writing \texttt{CMake}-code for the project, each
subdirectory has its own \texttt{CMakeLists.txt} file hence creating a
tree-structure with the main \texttt{CMakeLists.txt} in the root of the project
repository, while the ``leafs'' are in its respective subdirectories. However,
since \commander\ has a substantial codebase already, we have decided to split
the \texttt{Fortran} and \texttt{CMake} codes entirely. This allows for better
navigation, makes the directory structure cleaner, and thus easier to maintain
and expand. Furthermore, since the number of files involved in compiling
\commander\ and its dependencies is significant, we have therefore adopted a
so-called ``out-of-source'' build approach, as recommended by the
\texttt{CMake} creators. In this organization, the source code and compiled
build files are stored in separate locations. In our case, the source code
itself is located inside the directory called \texttt{commander3}. In contrast,
the binary folder is called \texttt{build}\footnote{It is worth noting that
there is nothing special about this name, and the users can name the directory
where the \texttt{CMake} build files will be stored in whatever way they
prefer. However, the convention is to name it \texttt{build}.}, and it is
usually created by the user inside the \texttt{root} directory. One advantage
of out-of-source compilation is that the users can create whatever amount of
\texttt{build} folders they want/require; thus, the same source code tree can
be used to produce multiple binaries, corresponding, for instance, to different
debug flags or CPU architectures.

All in all, the \commander\ source code directory structure is visualized in
Fig.~\ref{fig:source}. In this figure, \texttt{root} represents the root folder
of the project, created by the original \texttt{git clone} command;
\texttt{cmake} contains all \texttt{CMake} related files; \texttt{commander3}
contains all \commander\ related files; \texttt{docs} contains instructions on
how to generate out-of-source documentation; \texttt{logo} contains \commander\
logos; \texttt{.gitignore} is a \texttt{git} version control file;
\texttt{CMakeLists.txt} is the top level \texttt{CMake} configuration file,
which serves as the starting point for the compilation process;
\texttt{Makefile} is a traditional-style \texttt{Makefile} that may be used to
compile \commanderthree\ without \texttt{CMake}; and \texttt{README.md}
describes the project on \texttt{GitHub}.

\subsubsection{\commander\ \texttt{CMake} workflow}
\label{sec:cmake-code-organization}

\begin{figure*}[t]
      \center
      \includegraphics[scale=0.75]{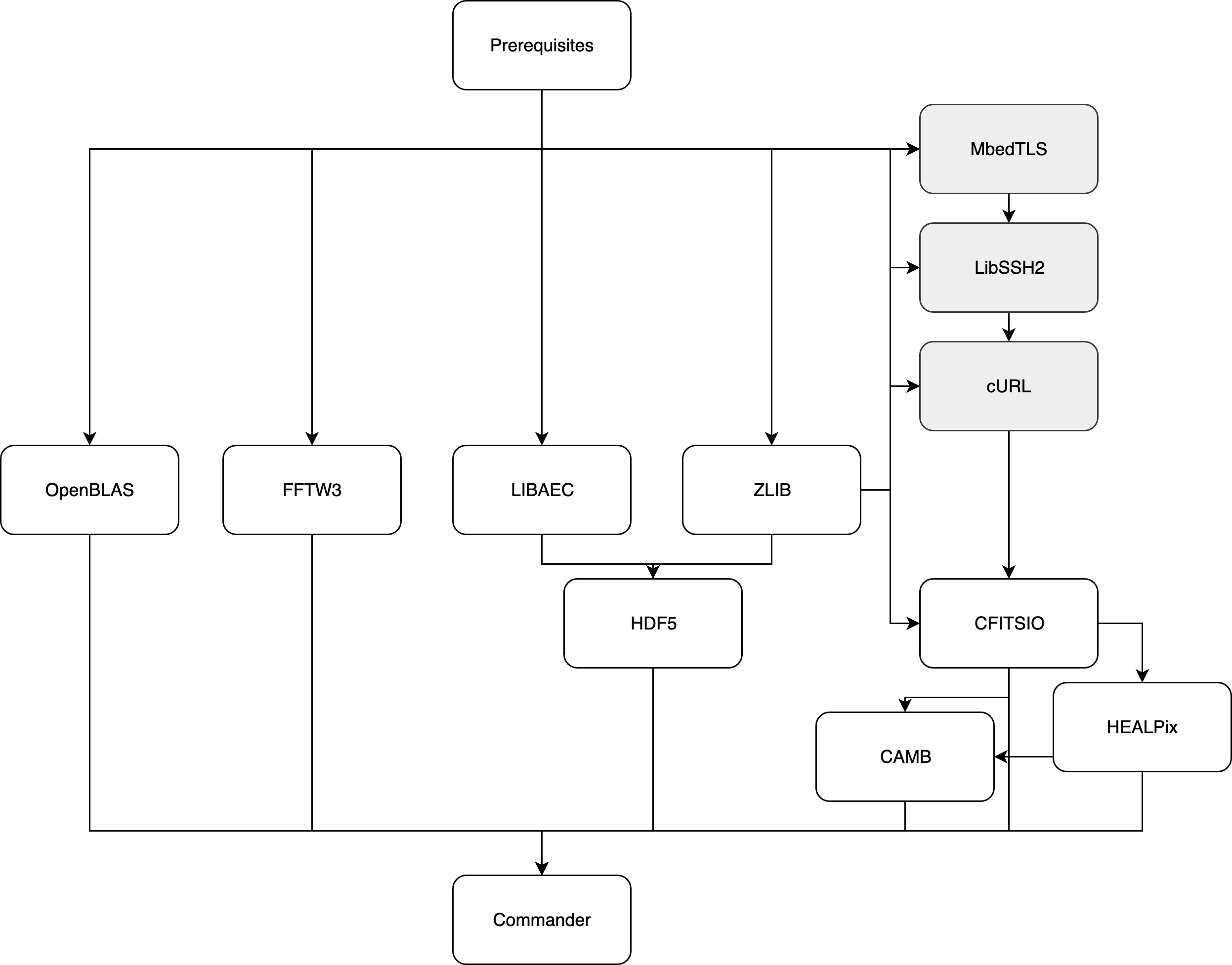}
      \caption{An example of the \texttt{CMake} \commander\ workflow. 
      The arrows represent the dependence and the linking, while the libraries 
      on the same level will be compiled in parallel. The libraries in grey are 
      optional dependencies. For example, \texttt{HDF5} is dependent on both \texttt{ZLIB} and \texttt{LIBAEC}. 
      The same goes for \texttt{CFITSIO} which may (or may not) be compiled with \texttt{cURL} 
      support depending on the user's preference. Since \texttt{cURL} is dependent on 
      a bunch of other libraries they also need to be incorporated into the build 
      even though \commander\ does not require them explicitly. 
      }
      \label{fig:cmake-workflow}
\end{figure*}

The \texttt{CMake} process works as follows. First, the host system is scanned,
and the present/missing  libraries are registered inside
\texttt{CMakeCache.txt} and other auto-produced files. Then, once the configure
step is done and the user has issued the build command, \texttt{CMake}
downloads, configures, compiles, and installs the missing dependencies and,
together with the ones identified as available, it links all libraries to the
compiled \commander.

The \texttt{CMake} module that enables this behavior is called
\texttt{ExternalProject}, and its primary purpose is to facilitate downloading
and installation of dependencies that are not an internal part of the main
project. In this way, the \commander\ dependencies are treated as entirely
independent entities. Such isolation allows the build to be performed in the
same way on different platforms, with utterly different build settings (e.g.,
compiler flags) and/or with a completely different build system (e.g.,
\texttt{Autotools}). Under the hood, it defines the set of so-called targets,
each representing a particular step in the build process of an external
project. These steps are then collected under a unified name (in our case, a
sub-project name), used later in the code. \texttt{CMake} also remembers
information about each performed step, which, if executed successfully, will
not be repeated. This allows us to compile all dependencies only once for
different \commander\ build types. An essential feature for debugging this
process (if and when something fails) is the \texttt{CMake} logs, which are
stored in \texttt{/usr/local/logs}\footnote{The location can be changed by
using defined \texttt{CMAKE\_LOG\_DIR} variable during the configuration
process. We refer the interested reader to \commander\ documentation for
further details.}.

Command-line arguments determine compiler selection during the scan phase. For
instance, \texttt{-DCMAKE\_Fortran\_COMPILER=ifort} tells \texttt{CMake} to use
the Intel \texttt{ifort} compiler. Then, for the most common compilers default
optimization flags are defined per (sub-)project in a configuration file called
\texttt{cmake/projects/{<project\_name>}.cmake}. When installing this software
on a new system with a new, unknown to \commander\, compiler, these are the
configuration files that most likely need to be updated.

Based on the initial system scan and user-specified compilation instructions,
\texttt{CMake} proceeds with the following steps for each dependency and for
the main \commander\ source (the two first steps are skipped in the latter
case):

\textbfit{Download:} The project is downloaded via external links in the form
of \texttt{.zip} or \texttt{.tar.gz} archives, or directly from \texttt{Git}
repositories. We use \texttt{MD5} hashes whenever possible to ensure that the
correct library versions are downloaded.

\textbfit{Update/Patch:} This step applies potential patches to the downloaded
archive or, in the case of \texttt{Git}, brings the project up to date. In
cases where we download the release versions of the packages, we skip this
step.

\textbfit{Configure:} This step can use \texttt{CMake} and other build tools
alike, depending on the preferences of the authors of the original dependency.
In our case, most libraries use \texttt{Autoconfig} scripts and
\texttt{Makefile} to compile.

\textbfit{Build:} In this stage, we use the default build tool as in the rest
of the project.

\textbfit{Install:} The subproject is installed to a local directory specified
by the user during the \texttt{CMake} configuration stage. It is worth noting
that not all projects (e.g., \texttt{HEALPix}) support an explicit install
command. We simply copy the compiled binaries and libraries into the specified
directory in these cases.

These steps can either be sequential or parallel. In the former case,  each
library is built sequentially, while the latter allows some libraries to be
built in parallel. This idea is illustrated in  Fig.~\ref{fig:cmake-workflow}.
It is important to note that this is not the only the way to compile \commander
since \texttt{OpenBLAS}  can be substituted for, e.g., Intel \texttt{MKL},
which will be detected by \texttt{CMake} if present. 

\subsubsection{Installation regimes}
\label{cmake-installation-regimes}

\texttt{CMake} has various build types defined by default that allows for
different optimization categories. We are calling these ``installation
regimes'' with four of these currently supported:

\texttt{Release:} Builds \commander\ with the most aggressive optimization
flags enabled, tuned for each specific compiler and platform. At the time of
writing this paper, only Intel and GNU compilers were supported.

\texttt{Debug:} Builds the \commander\ executable without any optimization, but
with debug symbols.

\texttt{RelWithDebInfo:} (``Release With Debug Information''). A compromise
between the two above, building the \commander\ binary with less aggressive
optimizations and with debug symbols.

\texttt{MinSizeRel:} (``Minimal Size Release''). Builds the \commander\
executable with optimizations that do not increase object code size. However,
as the current software targets HPCs with ample disk space, this feature is not
used frequently, and it is also accordingly not thoroughly tested.

While we have defined \texttt{RelWithDebInfo} to be the default one, the
installation regime is determined by the user and his/her needs and can be
changed via specifying \texttt{CMAKE\_BUILD\_TYPE} variable in the command
line. In general, all external libraries are produced in  \texttt{Release}
format with all optimization flags enabled. This can, however, be changed on a
case-by-case basis by editing the \texttt{CMake}  source files mentioned above.

\section{Documentation, QuickStart guide, and accessibility tools}
\label{sec:bp_reproducibility}

For an Open Science project to succeed and continue to grow, making the source
code and the data open to the general public is not sufficient. It is also
critical that the framework is easy to use and adapt from the user’s
perspective. Therefore, in this section, we provide a QuickStart guide to the
\commander\ documentation and compilation, as well as discuss some valuable
tools that make it easy for new users to get “up-to-speed” quickly. We note,
however, that this is a continuous work-in-progress; thus, the section is only
intended to give a snapshot of the situation at the time of publication.

While the \cosmoglobe\ framework is designed to simultaneously handle data from
different experiments, it is helpful to consider the specific \BP\
pre-processing and analysis pipeline in greater detail, as this will typically
serve as a point of reference for most users, whether they want to reproduce
the LFI work or generalize the framework to other datasets. In the following,
we, therefore, give an overview of the \BP\ installation procedure, but note
that most of these steps will be identical for any \commander-based analysis.

\subsection{Online documentation}
\label{sec:docs}

The documentation for the \BP\ pipeline and \commander\ are currently available
on the \commander\ documentation page of the official \cosmoglobe\
\texttt{GitHub} repository\footnote{\url{https://github.com/Cosmoglobe}}. It
consists of five main sections, namely (1) Overview (2) QuickStart and
installation guide; (3) The \commander\ parameter file; (4) File formats; and
(5) Frequently asked questions.

Due to the dynamic nature of the \commander\ and \cosmoglobe\ projects, this
documentation will continually evolve with the addition of new features or
experimental datasets. Hence, active participation by the community is
essential for its maintenance and expansion. Relatedly, it is also worth noting
that because of the high \commander\ development rate, sometimes developers
forget to document newly added parameters, and the documentation then becomes
outdated. When this happens, the code may request parameters that are not
explicitly mentioned anywhere. In these cases, we strongly recommend that the
external user notifies the core developers by opening a \texttt{GitHub} issue
-- or, better yet, corrects the documentation and submits the improved version
in the form of a \texttt{pull request}.

\subsection{\BP\ QuickStart guide}
\label{sec:quickstart}

In the ideal case, installing the \BP\ analysis framework can be done
in four simple steps:
{\small
\begin{verbatim}
> git clone https://github.com/Cosmoglobe/Commander.git 
> mkdir Commander/build && cd Commander/build
> cmake -DCMAKE_INSTALL_PREFIX=$HOME/local \
        -DCMAKE_C_COMPILER=icc \
        -DCMAKE_CXX_COMPILER=icpc \
        -DCMAKE_Fortran_COMPILER=ifort \
        -DMPI_C_COMPILER=mpiicc \
        -DMPI_CXX_COMPILER=mpiicpc \
        -DMPI_Fortran_COMPILER=mpiifort \
        ..
> cmake --build . --target install -j 8
\end{verbatim}}
\noindent The first line downloads the \commander\ from the official
\texttt{Git} repository, while the second line creates the aforementioned
\texttt{build} directory in which compiled binaries and libraries will be
stored. The third line collects information regarding the system and
auto-generates \texttt{Makefile}s inside \texttt{build}. Finally, the last line
downloads any potentially missing external dependencies and compiles both these
and the main source code\footnote{After installing \commander, the user needs
to update the shell to point out the location of the recently installed
\texttt{HEALPix} library. This is done by \texttt{export}ing the
\texttt{HEALPIX} variable inside \texttt{.bashrc} (or similar file).}. The
whole process typically takes less than 10 minutes. We have used Intel
compilers in this example, but GNU ones have also been successfully tested.

\begin{figure*}
      \center
      \includegraphics[width=18.5cm]{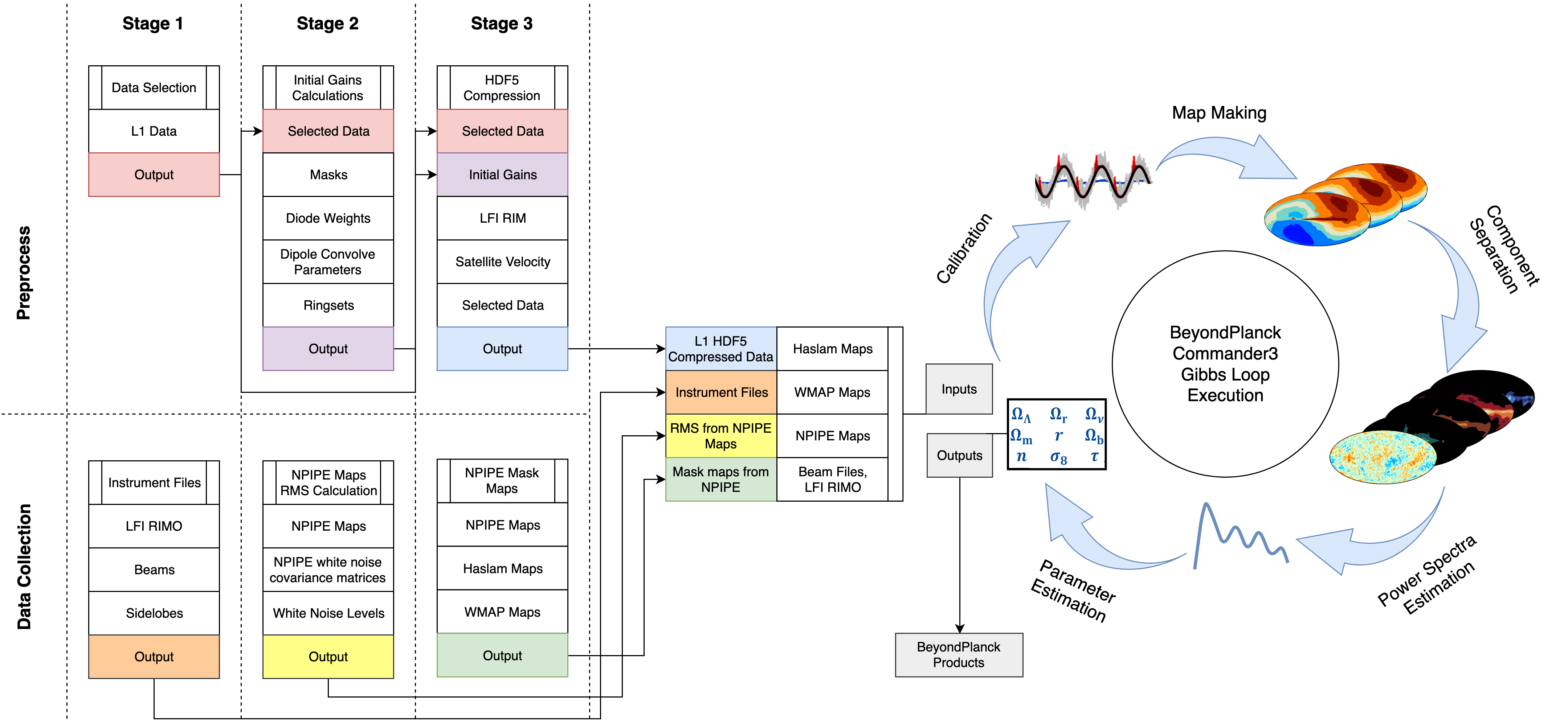}
      \caption{Schematic overview of the \BP\ pipeline pre-processing
        and initialization stages, along with all the input files
        required for each particular stage.}\label{fig:bp_pipeline_stages}
\end{figure*}

Once \commander\ is installed, the second step is to download the required
input data. The number of different files required for a complete \BP\ run can
be somewhat intimidating at first sight. To solve the problem, we have
implemented a small \texttt{Python} utility called \texttt{bp} that helps new
users to download all required data with a single command:

\begin{verbatim}
> bp download all
\end{verbatim}
\noindent This creates a complete directory structure with all required inputs,
which amounts to more than 1\,TB of data. With a modest download speed of
10\,MB/s, this can take some time before completion. The tool also supports
downloading individual sub-directories in case the user only requires a subset
of the total data.

The third step is to edit the \commander\ parameter file. As described by
\citet{bp03} and in the \commander\ documentation, this is a human-readable
ASCII file. It is the step with the steepest learning curve in the process, as
the number of \commander\ parameters is quite significant, and a typical
parameter file spans several thousands of lines. To address the issue, we have
enabled support for the nested include statements allowing for rarely used
parameters to be hidden from the user. The downside of this approach is that
the special environment variable should be defined in the user’s
shell\footnote{It is called \texttt{COMMANDER\_PARAMS\_DEFAULT}, and it should
point to the \texttt{<commander\_root>/commander3/parameter\_files/defaults}
inside \texttt{.bashrc} (or other shell files if applicable.)} for everything
to work. Even more so, although very helpful in most cases, such abstraction
can still lead to difficult-to-debug errors and become a potential time spender
since it requires considerable experience to debug \commander\ parameter files
efficiently. A good strategy, in this case, is to start with a well-tested case
(such as the final \BP\ parameter file) and only make a few changes between
each test run, carefully visually inspecting all outputs at each step while
gradually building intuition regarding the code outputs.

The fourth and final step is to actually run the code, which is typically done
through an \texttt{MPI} runtime environment:
\begin{verbatim}
> mpirun -n {ncore} path/to/commander param.txt
\end{verbatim}
The runtime for a given job varies wildly depending on the parameter file and
computing facilities. Still, for the default \BP\ parameter file and a 128-core
cluster, it takes about 1 hour and 40 minutes to produce one single sample
\citep{bp03}. For a full Monte Carlo posterior exploration that requires
thousands of samples, the end-to-end wall-time is typically on the order of
months.

This QuickStart guide represents the ideal case where everything works out of
the box. At the time of writing this paper, we estimated that the framework had
been successfully installed on at least 20 independent computer systems -- and,
unfortunately, the default process outlined above worked without modifications
in no more than half of these. In the remaining cases, various issues popped up
because of compiler idiosyncrasies, missing (or wrong version of) system
utilities, insufficient user permissions or disk space, etc. To solve such
issues when they arise and improve the current tools, it is vital to have a
deeper understanding of all parts of the process, which is the main topic of
the rest of the section.

\subsection{\BP\ pre-processing and initialization}
\label{sec:pipeline}

To understand the whole \BP\ analysis process, it is helpful to take a
high-level look at the entire pipeline. This is schematically illustrated in
Fig.~\ref{fig:bp_pipeline_stages}. The heart of this pipeline is the
\commanderthree\ execution, discussed in the previous section and illustrated
here by the rightmost analysis loop. This is where the actual posterior
sampling takes place (see Sect.~\ref{sec:posterior}), and it is implemented in
terms of a $\sim$\,60\,000 line \texttt{Fortran} code, as described by
\citet{bp03}.

However, \commander\ requires a significant number of input data objects in
order to run, as illustrated by the various small boxes to the left in the
figure. These include (1) the raw \Planck\ LFI Level-1 data (light blue box;
\citealp{planck2016-l02}); (2) a so-called ``instrument file'' (orange box);
(3) external and ancillary data that need no pre-processing (white boxes); and
(4) external or ancillary data that do need slight preprocessing to conform
with \commander\ conventions (colored boxes).

Going through these in order of low to high complexity, the white boxes
represent external sky maps and ancillary data that may be used directly in
their original forms. These include the frequency maps from \Planck\ HFI,
\WMAP, and Haslam 408\,MHz, and various instrument characterization such as
beam files and the \Planck\ LFI Reduced Instrument Model (RIMO). In many cases,
these may be simply downloaded directly from external repositories, such as the
\Planck\ Legacy Archive\footnote{\url{https://pla.esac.esa.int/}} or
LAMBDA\footnote{\url{https://lambda.gsfc.nasa.gov}}, and inserted into
\commander\ in their original form.

However, some information needs to be slightly pre-processed to match the
format expected by the \commander. One example is the white noise specification
per frequency band (yellow boxes), for which \commander\ expects the user to
provide a standard deviation per pixel, whereas the official \Planck\ products
provide a per-pixel $3\times 3$ covariance matrix. As such, the external user
needs to reformat the \Planck\ format into the \commander\ format (or, better
yet, implement support for the native \Planck\ format directly into \commander,
and submit a \texttt{Git} pull request).

Another important example is masks (green boxes), which are used at various
stages during the \commander\ processing. These may be defined differently
whether one is considering correlated noise, gain, bandpass, foreground, or CMB
estimation \citep[e.g.,][]{bp06,bp07,bp11,bp12,bp13,bp14}. These masks are
typically based both on external sky maps (e.g., \Planck\ HFI or \WMAP) and
internal results from a previous \commander\ iteration (e.g., $\chi^2$ maps),
and properly optimizing these is an important (and non-trivial) task for any
\commander\ user.

The third data collection box represents the so-called \commander\ instrument
file. This plays a similar role as the RIMO in \Planck and contains detailed
instrument information for a given frequency channel and detector. This
includes objects that are general for all detectors, such as bandpasses and
beams \citep{bp08,bp09}, but also instrument-specific objects such as ADC
correction tables \citep{bp25}.

However, the most significant and crucial pre-processing step is the
preparation of the actual raw time-ordered data, as indicated by the three
``Preprocess'' stages. These data are stored in compressed \texttt{HDF5} files
\citep{bp03}, and include everything from raw detector readouts, pointing,
flags, and satellite velocity to initial gain and noise estimates per \Planck\
pointing period.

To improve both user-friendliness and reproducibility in each of the above
steps, it is useful to define scripts that perform all these tasks for the
user. Within the \commander\ repository, we have therefore provided a series of
(primarily \texttt{Python}) scripts that perform each of these operations, from
mask and instrument file generation to full Level-1 data processing. These are
intended to serve as useful starting points for users who seek to reproduce the
current \BP\ LFI processing and for users who want to analyze a completely new
dataset with the same framework. If so, it would be greatly appreciated if the
new scripts are also committed to the existing repository as part of the
community-driven Open Source activities.

Before concluding this section, it is worth noting that \commander\ has, in
general, very few means of validating a given input data product. If, say, some
instrument specification or compressed \texttt{HDF} files are inter-mixed,
there is no automatic way for the algorithm to discover this except through
visual inspection of the final results and goodness-of-fit statistics.
Furthermore, such parameter file errors are likely among the most common and
time-consuming errors made when running this code. For most analyses, it is,
therefore, useful to start with the set of well-tested parameter and input
files provided in the \commander\ repository \citep{bp03}, which includes
individual parameter files for a wide range of common datasets (\Planck\ LFI
and HFI, \WMAP, Haslam 408\,MHz, etc.) and astrophysical components (CMB,
synchrotron, thermal dust emission, etc.). These may be used as ``building
blocks'' when constructing a new analysis configuration.

\subsection{Docker environment for user-friendly data access and code exploration}

We provide a precompiled \texttt{Ubuntu}-based \texttt{Docker} image for users
who are not interested in computationally expensive analyses like \BP\ but
simply want to run \commander\ on a small dataset for which computational
efficiency is not paramount. This self-contained operating-system-level virtual
container can be run on any OS (\texttt{Mac}, \texttt{Windows}, \texttt{Linux},
etc.), and all dependencies are maintained within the \texttt{Docker} image
itself. Running \commander\ in this mode amounts to one single command line:
{\small
\begin{verbatim}
> docker run -it \
  -v {input_dir}:{input_dir} \
  -v {output_dir}:/output \

  registry.gitlab.com/beyondplanck/r13y-helper/cm3 \
  commander3 {parameter file}
\end{verbatim}
}
\noindent where \texttt{input\_dir} is a directory that contains all required
input data, and \texttt{output\_dir} is an empty directory that will
contain the results. 

We emphasize, however, that the binary provided in this Docker image is not
optimized for any processor type. Therefore, it is computationally less
efficient than a natively compiled version. Also, debugging this version is
non-trivial since it may be challenging to recompile the binary with different
debug flags or source code changes.

\section{Discussion}
\label{sec:conclusions}

The successes achieved in modern CMB cosmology during the last decades are a
solid testament to the ingenuity and dedication of thousands of
instrumentalists, observers, data analysts, and theorists. However, these same
successes are also a direct product of long-standing and invaluable financial
support from ordinary taxpayers. A typical CMB satellite mission costs several
hundreds of millions of dollars, euros, or yen. At the same time, ground-based
and sub-orbital experiments typically cost from a few to many tens of millions
of dollars -- and the massive next-generation ground-based CMB-S4 experiment is
anticipated to cost \$\,600\,M.

With steadily rising costs for each generation of experiments, it also becomes
even more critical to optimally leverage the investments already made from
previous efforts. For instance, it makes very little sense for future
experiments to reproduce the temperature sensitivity of \Planck. Instead, they
should aim to provide complementary information that may be combined with the
\Planck\ measurements, typically in polarization or on small angular scales.
Likewise, it makes very little sense for a future satellite mission, such as
\textit{LiteBIRD}, to measure small angular scales from space when this can be
done much more economically from the ground at a much lower cost, for instance,
with CMB-S4.

In this paper, we argue that the most efficient way to move forward as a field
is precisely through an integrated joint global analysis of complementary
datasets. However, several prerequisites must be in place for this to be
possible. First and foremost, researchers in the various teams actually need to
have physical access to data from other experiments. Traditionally, this has
been achieved through dedicated ``Memoranda of Understanding'' (MoUs) between
pairs of collaborations; the ground-breaking joint analysis of \Planck\ and
BICEP2 is a well-known example of this \citep{pb2015}. While this works
reasonably well for limited two-party cases, we believe that this approach is
impractical for future work when many datasets must be involved in the same
analysis to obtain optimal results, for instance, when combining proprietary
data from \textit{LiteBIRD} \citep{litebird2020}, CMB-S4 \citep{cmbS4}, C-BASS
\citep{cbass18}, QUIJOTE \citep{QUIJOTE_I_2015}, and PASIPHAE
\citep{tassis:2018} with public data from \Planck\ and \WMAP. Rather, we
believe the time is overdue to fundamentally change how the CMB field works and
move to a fully Open Science mode of operation where both raw data and
end-to-end analysis methods are shared between experiments and research groups.
This also guarantees that taxpayers and funding agencies get maximum value for
money, as it vastly increases the longevity of any given dataset. Hence, the
funding agencies should require an Open Source release of raw data, analysis
tools, and high-level products for any future experiment.

A second prerequisite is using practical analysis methods to exploit the
information stored in these complementary datasets. Establishing one set of
such common tools is the primary goal of the \BP\ and \cosmoglobe\ efforts. Our
implementation is based on well-established Bayesian parameter estimation
techniques and builds directly on the \commander\ code developed for and used
by \Planck. This software is released under a permissive Open Source GPL
license, ensuring that any researchers may use, generalize, and modify the code
as they see fit \citep{bp01,bp03}. The only requirement is that these modified
versions must also be released under an equally permissive license, ensuring
that other scientists may then also benefit from the extended work. 

However, it is still not sufficient that the data and analysis codes are
publicly available. They must also be \emph{accessible}, and that is the main
topic of the current paper: For other scientists to be able to leverage this
work in practice, the software must be appropriately documented, and it must be
straightforward to install on a range of different computer systems. These
practical aspects may seem somewhat mundane compared to the more spectacular
topics typically addressed in astrophysics and cosmology papers – but they are
no less important in this era of mega-science. These aspects also require
significant dedicated resources to be successful, and \BP\ dedicated as much as
20\,\% of its budget (or 300\,k\texteuro) to this work. On the other hand,
these resources do not scale linearly with the total budget size. However, we
still strongly recommend future experiments allocate significant funding for
reproducibility and Open Source dissemination in their proposal budgets. We
hope that the current paper may serve as a valuable and thought-provoking
starting point for future large experiments and collaborations that are likely
to face similar issues.


\section*{Acknowledgements}
{\small
  We thank Prof.\ Pedro Ferreira and Dr.\ Charles Lawrence 
  for useful suggestions, comments and 
  discussions. We also thank the entire \Planck\ and \WMAP\ teams for
  invaluable support and discussions, and for their dedicated efforts
  through several decades without which this work would not be
  possible. The current work has received funding from the European
  Union’s Horizon 2020 research and innovation programme under grant
  agreement numbers 776282 (COMPET-4; \BP), 772253 (ERC;
  \textsc{bits2cosmology}), and 819478 (ERC; \textsc{Cosmoglobe}). In
  addition, the collaboration acknowledges support from ESA; ASI and
  INAF (Italy); NASA and DoE (USA); Tekes, Academy of Finland (grant
   no.\ 295113), CSC, and Magnus Ehrnrooth foundation (Finland); RCN
  (Norway; grant nos.\ 263011, 274990); and PRACE (EU). Some of the results 
  in this paper have been derived using the \texttt{healpy} \citep{zonca2019} 
  and \texttt{HEALPix} \citep{gorski2005} packages.
}

\bibliographystyle{aa}
\bibliography{./common/Planck_bib, ./common/BP_bibliography}

\end{document}